\documentclass{WileyMSP-template}

\usepackage{float}
\usepackage{verbatim}
\usepackage[utf8]{inputenc}
\usepackage[T1]{fontenc}
\usepackage[english]{babel}
\usepackage{dsfont}
\usepackage[normalem]{ulem}
\usepackage{amsmath}
\usepackage{amssymb}
\usepackage{mathrsfs}
\usepackage{amsthm}
\usepackage{mathtools}
\usepackage{wasysym}
\usepackage{breqn}
\usepackage{comment}
\usepackage{bbm}
\usepackage{nicefrac}
\usepackage{soul}
\usepackage{stackrel}
\usepackage{braket}
\usepackage{enumitem}
\usepackage{graphicx}
\usepackage{tikz}
\usepackage{mdframed}
\usepackage{booktabs}
\usepackage{siunitx}
\usepackage{comment}
\usepackage{subcaption}
\usepackage[symbol]{footmisc}

\usepackage{calc}
\usepackage{accents}
\newcommand{\dbtilde}[1]{\tilde{\raisebox{0pt}[0.85\height]{$\tilde{#1}$}}}

\usepackage{multirow}
\usepackage{array}
\newcolumntype{P}[1]{>{\centering\arraybackslash}p{#1}}

\usepackage[hidelinks]{hyperref}
\usepackage{framed}

\newcommand{\ketbra}[2]{\ket{#1}\bra{#2}}

\addto\captionsenglish{}
\newcommand{\norm}[1]{\left\|#1\right\|}

\DeclareMathOperator{\Tr}{Tr}

\newcommand{\one}{\mathds{1}}

\newcommand{\cor}[1]{#1}

\usepackage{cases}

\addto\captionsenglish{}

\usepackage{algorithm}

\makeatletter
\newenvironment{protocol}
{
		\renewcommand{\ALG@name}{Protocol}
		\refstepcounter{algorithm}
		\hrule height.8pt depth0pt \kern2pt
		\renewcommand{\caption}[2][\relax]{
			{\raggedright\textbf{\fname@algorithm~\thealgorithm} ##2\par}%
			\ifx\relax##1\relax 
			\addcontentsline{loa}{algorithm}{\protect\numberline{\thealgorithm}##2}%
			\else 
			\addcontentsline{loa}{algorithm}{\protect\numberline{\thealgorithm}##1}%
			\fi
			\kern2pt\hrule\kern2pt
		}
	}{
	\kern2pt\hrule\relax
}
\makeatother

\makeatletter
\let\cat@comma@active\@empty
\makeatother

\begin{document}

\pagestyle{fancy}
\rhead{\includegraphics[width=2.5cm]{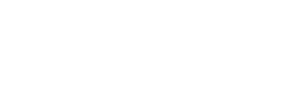}}

\title{Efficient simulation of noisy entanglement generation}

\maketitle


\author{Lorenzo Brevi}

\author{Federico Grasselli\footnote[1]{corresponding author: federico.grasselli@leonardo.com}}

\author{Alessandro Caraceni}

\author{Massimiliano Proietti}

\author{Massimiliano Dispenza}

\author{Enrico Prati\footnote[1]{corresponding author: enrico.prati@unimi.it}}


\dedication{}

\begin{affiliations}
Lorenzo Brevi, Prof. Enrico Prati\\
Department of Physics Aldo Pontremoli, Università degli Studi di Milano, Via Celoria 16, 20133 Milano, Italy
\\
Istituto di Fotonica e Nanotecnologie, Consiglio Nazionale delle Ricerche, Piazza Leonardo da Vinci 32, 20133 Milano, Italy

Dr. Federico Grasselli, Alessandro Caraceni, Dr. Massimiliano Proietti, Dr. Massimiliano Dispenza\\
Leonardo Innovation Labs -- Quantum Technologies, Via Tiburtina km 12400, 00131 Rome, Italy

\end{affiliations}



\begin{abstract}
    End-to-end entanglement distribution is a key capability of upcoming quantum networks, enabling applications like distributed quantum computing, quantum sensor networks, and secure communications. Hence, its realistic and efficient simulation is crucial for quantum network design and for assessing the ability of a network to run certain applications. This work provides tools to scale-up and improve the realism of entanglement generation simulations in quantum networks. This is achieved by deriving analytical results that directly return the success probability, the output state and corresponding fidelity of a selected entanglement generation protocol, while accounting for a variety of noise sources affecting the protocol. These results are then integrated and streamlined in an upgraded version of SeQUeNCe, one of the most popular quantum network simulators. The resulting simulator features increased scalability by reducing computation time by more than $60 \%$, while allowing for a variety of realistic noise sources, including imperfect mode matching, dark counts, and imperfect memory initialization. The simulator is also benchmarked with real experimental data and is capable of replicating the average entanglement generation time and the final state fidelity of a selected experiment.
    Altogether, the results can enhance current quantum network simulation capabilities towards large-scale networks, paving the way for the future quantum internet.

\end{abstract}



\section{Introduction}

Similarly to how Internet had a revolutionary impact on our world, a quantum internet \cite{doi:10.1126/science.aam9288, qunatum_int_protocol}, that is, a global infrastructure interconnecting quantum networks for the reliable transmission of quantum information and distribution of entanglement, can lead to groundbreaking applications \cite{Federico_review}. These include, for instance, enhanced sensing capabilities by linking sensor networks \cite{zang2025quantumadvantagedistributedsensing}, scaling-up quantum computing power via distributed quantum computing \cite{nature_distributed, distr_1,cuomo2023optimized},  cryptographic services with information-theoretic security \cite{cryptography1, cryptography2,cavaliere2020secure},  atomic clocks synchronization \cite{common-loss}, and blind computation \cite{blind1}.

While a large-scale quantum network is still out of technological reach \cite{exp1, exp2, exp3, exp4, exp5, exp7, exp8, exp9}, an important prerequisite for realizing a quantum internet is the ability to digitally simulate quantum networks \cite{NetSquid,Satoh_2022,SimQN,SeQUeNCe}. A simulator is required in order to gauge the characteristics of larger quantum networks that cannot be fully modeled analytically \cite{NetSquid,Vardoyan_2021,Yehia_2024}. However, even in small networks, the interactions among multiple communicating devices driven by their timing dependencies make it hard to analytically characterize the performance metrics. Therefore, simulators become an essential tool to guide the design of a future quantum internet. One of the most prominent quantum network simulators is the Simulator of QUantum Network Communication (SeQUeNCe) \cite{SeQUeNCe}. SeQUeNCe is an open-source simulator that is extremely customizable due to its modular nature and is capable of simulating hybrid classical and quantum networks \cite{comp}.

One of the most important primitives of quantum networks -- and of quantum network simulators -- is the generation of entanglement between two quantum memories hosted in neighboring nodes, as it represents the first necessary step to distribute end-to-end entanglement across the network. There exist a variety of protocols to establish entanglement between two quantum memories. In this work, we focus on the Barrett-Kok (BK) protocol \cite{Barrett-Kok}, which heralds the generation of entanglement between two quantum memories by performing Bell state measurements and it is also used in SeQUeNCe. In the real world, entanglement generation is affected by many noise sources that can greatly impact both the time required to successfully establish entanglement and its quality. However, in several current simulators, including SeQUeNCe, important error sources are not properly modeled, and, when so, they often introduce inefficiencies and delays in the simulations.

Here, we lay the foundation for more realistic and efficient simulations of  entanglement generation in quantum network simulators. The progress is enabled by the derivation of generalized analytical expressions for the output state of the BK protocol, its fidelity and the success probability of the protocol, while taking into account the effects of a variety of noise sources. The analytical results following such a generalization are then employed to upgrade the quantum network simulator SeQUeNCe, allowing it to account for additional noise sources and simultaneously streamlining its computational pipeline. Altogether, our results improve both the realism and scalability of the simulator, providing a robust foundation for future quantum internet simulations.

More in detail, the entanglement generation routine in our version of SeQUeNCe accounts for memory and detector efficiency, channel transmittance, imperfect state preparation (i.e., the initialization of the memory state), photon distinguishability and detector dark counts. The derived analytical expressions allow us to significantly reduce the number of computations performed by the simulator in each entanglement generation attempt, compared to simulating the effect of each noise source with a sequence of gates applied on the initial state of the memories. As a result, our simulator includes several error sources of entanglement generation while being $60 \%$ faster then the original version of SeQUeNCe. We then apply our simulator to replicate a real experiment of entanglement generation between two nitrogen–vacancy centers \cite{exp_3m} and obtain a value for the fidelity within one standard deviation of the experimental result. Furthermore, our simulator is able to replicate the distribution of protocol attempts before a successful entanglement, thus obtaining an estimate of the required time in line with the experimental result from \cite{exp_3m}.

The work is organized as follows. In Section~\ref{sec:noisy_entanglement} we first illustrate the Barrett-Kok protocol (Subsection~\ref{sec:protocol}), next we list the noise sources we account for (Subsection~\ref{sec:noise_sources}) and report the main analytical expressions obtained from our analytical model, i.e. success probability, output state, fidelity (Subsection~\ref{sec:analytical_model}). We turn to the description of the steps to obtain our analytical expressions (Subsection~\ref{sec:simplified_simul}). We therefore utilize these results to upgrade SeQUeNCe in Section \ref{sec: simul_seq}. The main changes in the code are explained in Subsection \ref{sec: framework} and compared with the native version of SeQUeNCe in Subsection \ref{sec: comp}. We then employ our simulator to study the effects of the error parameters on entanglement generation time and state fidelity in Subsection \ref{sec: simul}. Lastly, in Subsection \ref{sec: exp} our findings are benchmarked with a real experiment from the literature. We draw conclusions in Section \ref{sec:conclusions}. In Appendix~\ref{app:calculations} we report the complete derivation of our analytical expressions when all noise sources are combined. In Appendix~\ref{sec: photon_ind} we show how photon indistinguishability is accounted for in our analytical model.


\section{Noisy entanglement generation}\label{sec:noisy_entanglement}
    
\begin{figure}[htb]
    \centering
    \includegraphics[width=0.7\linewidth]{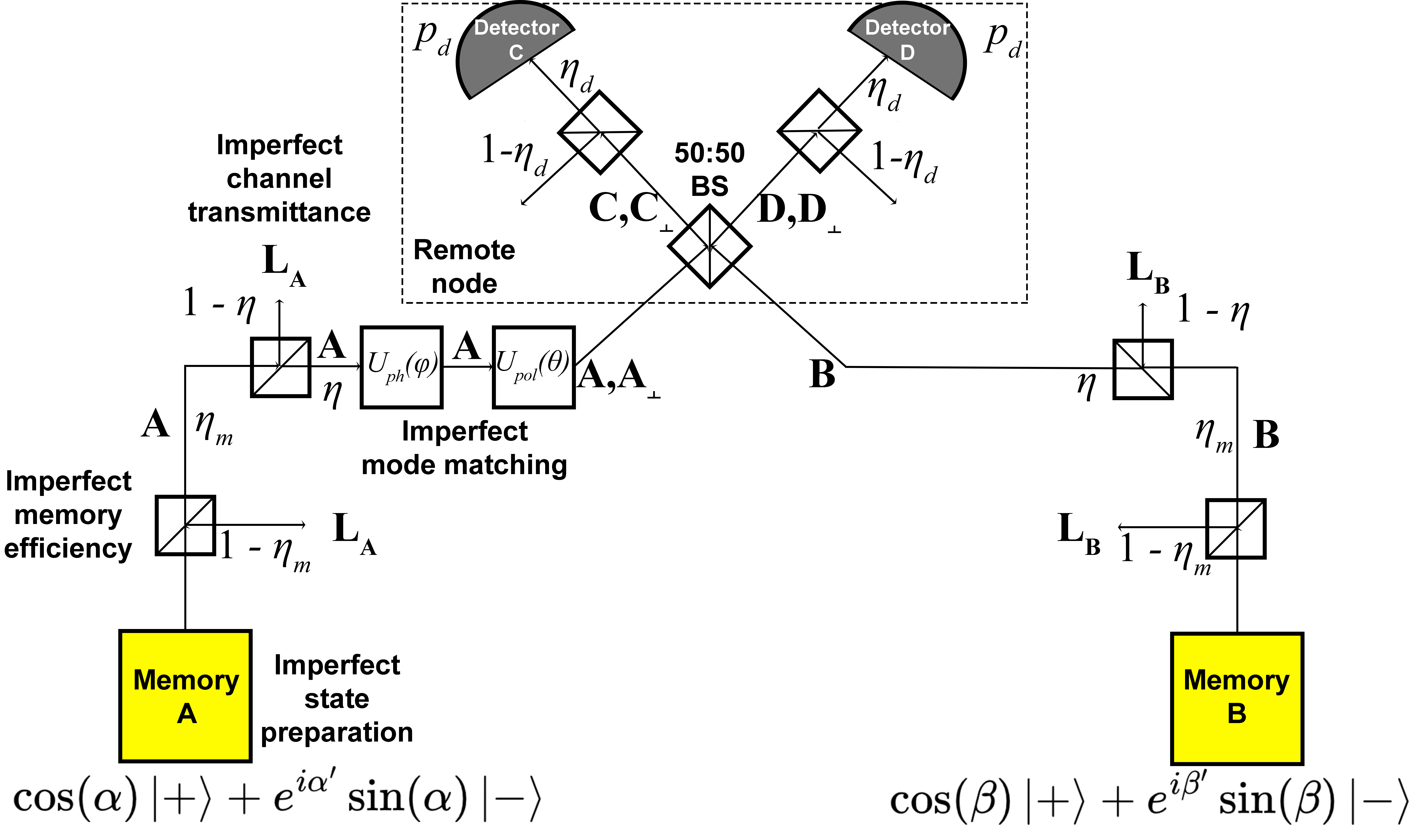}
    \caption{The optical setup required by the Barrett-Kok (BK) protocol, which enables two spatially-separated quantum memories, A and B, to probabilistically establish an entangled state. The setup depicts the way in which we modeled the various error sources accounted for in this work.
    The memories are initialized imperfectly in the states  $\cos\alpha \ket{+} + e^{i\alpha'} \sin\alpha \ket{-}$ for memory A and $\cos\beta \ket{+} + e^{i\beta'} \sin\beta \ket{-}$ for memory B, where $\ket{+} = (\ket{\uparrow} + \ket{\downarrow})/\sqrt{2}$ represents the ideal initial state of the memory in the BK protocol. Upon excitation of the $\ket{\downarrow}$ state, each memory releases a photon which is coupled in the optical channel with probability $\eta_m$. The photons traverse a quantum channel with probability $\eta$ before being combined in a balanced beam splitter followed by two threshold detectors with efficiency $\eta_d$ and dark count probability $p_d$. All photon losses are modeled by beam splitters with appropriate transmittance. Moreover, the photons from the two memories are not fully indistinguishable. This fact is modeled by applying on mode $A$ two unitaries, $U_{ph}(\varphi)$ and $U_{pol}(\theta)$, which introduce a phase mismatch of angle $\varphi$ and a non-zero component in an orthogonal mode (mode $A_\perp$).}
    \label{fig:BK-setup}
\end{figure}

Throughout the derivation, the protocol adopted for generating entanglement between two memories is the Barrett-Kok (BK) protocol and we implicitly refer to it when talking about entanglement generation. In this Section, we first describe the BK protocol as modeled in  the SeQUeNCe's white paper \cite{SeQUeNCe}. We report a scheme of the protocol in \textbf{Figure \ref{fig:BK-setup}}. Then, we list the noise sources accounted for in our calculations and present the success probability and final fidelity of the BK protocol. Finally, we illustrate the steps of the calculation in a simplified scenario with only a subset of noise sources (see Appendix~\ref{app:calculations} for the complete calculation).


\subsection{The Barrett-Kok protocol} \label{sec:protocol}

The entanglement setup of the BK protocol is summarized in Figure \ref{fig:BK-setup}. The goal is to entangle two distant emissive memories, A and B. The photons emitted by each memory are coupled to a quantum channel and transmitted to a central measurement node. The measurement node combines the light modes from the memories in a balanced beamsplitter and measures the outputs with a pair of threshold detectors, C and D. The detectors are also connected to memories A and B via a classical channel in order to communicate the results of the measurement. Each quantum memory can be prepared in two long-lived states, $\ket{\uparrow}$ and $\ket{\downarrow}$. The $\ket{\downarrow}$ state can be excited to another state $\ket{e}$ by applying an optical pulse of the right frequency. In a relatively short time frame, the state $\ket{e}$ then relaxes back to $\ket{\downarrow}$ while emitting a photon.
The BK protocol reads as follows.\\
\vspace{1ex}

\begin{protocol} \label{prot} \caption{BK protocol}
\begin{enumerate}[wide, labelwidth=!, labelindent=0pt]
\item Prepare each memory in the state $(\ket{\uparrow} + \ket{\downarrow})/\sqrt{2}$.

\item Apply an optical $\pi$-pulse to each memory, thereby inducing the transitions $\ket{\uparrow} \to  \ket{\uparrow}$ and $\ket{\downarrow} \to \ket{e}$, where $\ket{e}$ is an excited state.

\item Wait for a sufficiently long time for each memory to relax, inducing the transitions $\ket{e} \to \ket{\downarrow} \ket{1}$, where the state of the memory is now coupled to the emission of a photon. If, within this time interval, no detector clicked or both detectors clicked, the protocol aborts.

\item Conditioned on not aborting in the previous step, apply the Pauli $X$ gate on each memory qubit (spin flip).

\item Repeat steps 2. and 3.

\item If the two consecutive detections did not occur in the same detector, apply the Pauli $Z$ gate on the second quantum memory. If the protocol did not abort, the two memories are (ideally) in the Bell state $\ket{\Psi^+}=(\ket{\uparrow}\ket{\downarrow} + \ket{\downarrow}\ket{\uparrow})/\sqrt{2}$.
\end{enumerate}
\end{protocol}\vspace{2ex}

The BK protocol essentially boils down to two consecutive Bell state measurements in the central node. Each measurement relies on single-photon interference effects to post-select the state of the memories in an entangled state, upon detection in only one detector. Indeed, suppose that the left detector clicks in step 3 of the protocol, after the first excitation cycle. Due to the presence of the BS which eliminates the path information, this click can be caused by a photon emitted by either memory. That is, it could be that the left memory is in state $\ket{\downarrow}$ after emitting the photon while the right memory is in state $\ket{\uparrow}$ and did not emit any photon, or viceversa. The superposition of these two possibilities projects the state of the memories on the entangled state $\ket{\Psi^+}$. 

However, there is also the possibility that both memories emitted a photon and only one click was registered due to the Hong-Ou-Mandel effect \cite{HOM}, or due to photon loss. This implies that the state of the memories conditioned on a single detector click contains spurious elements such as $\ket{\downarrow}\ket{\downarrow}$. The BK protocol manages to remove such contributions by applying the Pauli $X$ before initiating the second excitation cycle, mapping such spurious contributions to $\ket{\uparrow}\ket{\uparrow}$ and thus preventing them from generating photons in the next excitation cycle. By conditioning again on a single detector click, the spurious contributions are removed from the joint state of the memories, which are post-selected in the Bell states $\ket{\Psi^{\pm}}$.

As a result, the BK protocol is naturally robust against losses. In Subsection~\ref{sec:simplified_simul}, we illustrate this explicitly by computing the final state produced by the BK protocol when only losses are present and show that it recovers the Bell states $\ket{\Psi^{\pm}}$. Note however, that the BK protocol is based on a linear optics setup and as such it remains a probabilistic entanglement generation protocol with success probability ideally bounded by $1/2$ \cite{max_BS}.

\subsection{Noise sources}  \label{sec:noise_sources}

Let's now turn to the noise sources accounted for in the following. While we acknowledge that some important sources of error are still missing, we believe that the errors considered here can already prove useful in correctly modeling an experimental setup implementing the BK protocol, as discussed in Subsection \ref{sec: exp}.

The noise sources accounted for in the derivation of our analytical results are the following:

\begin{enumerate}
    \item \textit{Photon losses}\\
    All sources of loss are modeled by applying a beam splitter (BS) on the photonic mode, with transmittance $\eta$ equal to the probability that one photon is not lost and where the other input is the vacuum. Formally, this is achieved by introducing a unitary $U_{loss}(\eta)$ which acts as follows on the creation operator of the affected photonic mode:
    \begin{align}
         a_{A(B)}^\dag \xrightarrow{U_{loss}(\eta)} \sqrt{\eta} \, a_{A(B)}^\dag + \sqrt{1-\eta} \, a_{L_A (L_B)}^\dag \label{loss-unitary},
    \end{align} where $L_A$ ($L_B$) is the mode of the lost photon.

    In this work, we consider three sources of loss: 1) the channel loss, modeled by a BS with transmittance $\eta$ placed in the channel between the memory and the interference measurement; 2) the detector efficiency given by $\eta_d <1$, modeled by a BS with transmittance $\eta_d$ placed before the detector; 3) the photon emission efficiency of the quantum memory\footnote{Note that, in practice, this efficiency is influenced by the emission of photons in modes other than the cavity mode, by the light leakage rate of the cavity and by the coupling to the fiber.}, given by $\eta_m<1$, modeled by a BS with transmittance $\eta_m$ placed before the quantum channel.

     \item \textit{Imperfect mode matching}\\
     Imperfect mode matching of the photons emitted by the memories reduces the fidelity of the output state, because the photons carry information regarding their origin thereby destroying interference. In particular, one could account for polarization rotations and phase shifts occurring in the quantum channels, while mismatching frequencies could be avoided by carefully tuning the parameters of each cavity \cite{Barrett-Kok}. In order to account for polarization mismatches of the interfering photons, we apply the following polarization rotation only on the photons in mode $A$, i.e. from the left memory, before they reach the balanced beam splitter:
     \begin{align}
         a_A^\dag \xrightarrow{U_{pol}(\theta)} \cos(\theta) \, a^\dag_{A} + \sin(\theta) \, a_{A_{\perp}}^\dag,   \label{imperfect_mode-unitary}
     \end{align}
     where $a_{A_{\perp}}^\dag$ creates photon in an orthogonal polarization mode. Similarly, to account for phase mismatches, we apply the following phase on the photons in mode $A$:
     \begin{align}
         a_A^\dag \xrightarrow{U_{ph}(\varphi)} e^{i\varphi}\, a_A^\dag.
     \end{align}

     We note, however, that the rotation in \eqref{imperfect_mode-unitary} can be interpreted as modeling a set of mode mismatches (including, e.g., imperfect overlap of the temporal packets) and not just polarization mismatches. In this vein, it is useful to connect this noise source with the indistinguishability $I$ of the photons, a parameter that is more commonly characterized in experimental setups. This is done by the relation:

     \begin{equation} \label{eq: iVStheta}
         I = \cos^2(\theta),
     \end{equation}
     obtained in Appendix \ref{sec: photon_ind}.

    \item \textit{Detector dark counts}\\
    Dark counts in the two detectors, in either of the two detection moments, can post-select the wrong state. However, their effect can be reduced by collecting clicks only in the intervals when the photons are expected to arrive. In order to account for dark counts, we replace the POVM element corresponding to a click in a detector, which is $\one - \ketbra{0}{0}$ in the ideal case, with: $\one - (1-p_d)\ketbra{0}{0}$, where $p_d$ is the dark count probability in one measurement round and is obtained by multiplying the dark count rate (dark counts per second) with the duration of the detection interval. Moreover, since the detectors are not assumed to distinguish the different modes of the impinging photons, the correct modeling of POVM element corresponding to a click is: $\one_C \otimes \one_{C_\perp } - (1-p_d)\ketbra{0}{0}_C \otimes \ketbra{0}{0}_{C_\perp }$.
    
    \item \textit{Imperfect state preparation}\\
    The state in which each memory is initialized might deviate from the balanced superposition assumed in the BK protocol, namely $(\ket{\uparrow} + \ket{\downarrow})/\sqrt{2}$. We model this by assuming that the initial state of the left memory reads:
    \begin{align} \label{eq: imperfect_state_prepA}
        \ket{\psi_1}_{M_A} = \cos\alpha \ket{+} + e^{i\alpha'} \sin\alpha \ket{-},
    \end{align}
    where we defined the superposition states: $\ket{\pm} = (\ket{\uparrow} \pm \ket{\downarrow})/\sqrt{2}$. Similarly, the initial state of the right memory reads:
    \begin{align} \label{eq: imperfect_state_prepB}
        \ket{\psi_1}_{M_B} = \cos\beta \ket{+} + e^{i\beta'} \sin\beta \ket{-}.
    \end{align}
\end{enumerate}

In addition to the above noise sources, the native version of SeQUeNCe accounts for the decoherence time of the memories. However, rather than continuously updating their state based on the passing of time, SeQUeNCe leaves the memory state unchanged and automatically resets it if the time elapsed from the last reset exceeds the memory decoherence time.  \cor{Furthermore, it is important to note that we do not need to take into account imperfect single qubit operations among the noise sources, as their effect is negligible when compared to the other sources described above.}

\subsection{Analytical model}  \label{sec:analytical_model}

By executing the steps of the BK protocol as outlined in Subsection~\ref{sec:protocol} while accounting for the realistic noise sources reported in Subsection~\ref{sec:noise_sources}, one can expect that the resulting success probability and fidelity of the output state deviate from the ideal ones. In Appendix~\ref{app:calculations}, we report the full derivation of the analytical expressions for the fidelity of the output state and for the success probability of the BK protocol. For brevity, here we only report the resulting expressions as a function of the noise parameters introduced in Subsection~\ref{sec:noise_sources}.

The success probability of the BK protocol is given by:
\begin{align}
    \Pr(\textrm{success}) 
    &=(1-p_d)^2 \left\lbrace p_d^2 (1-\eta_t)^2 (\tilde{x}_{\uparrow\uparrow} + \tilde{x}_{\downarrow\uparrow} + \tilde{x}_{\uparrow\downarrow} +\tilde{x}_{\downarrow\downarrow}) \right. \nonumber\\
    &\left.\quad + p_d \eta_t \left[(1-\eta_t) \left(\tilde{x}_{\downarrow\uparrow}  + \tilde{x}_{\uparrow\downarrow}  \right) + \left(1-\eta_t\left(\frac{3}{4}-\frac{\cos^2\theta}{4}\right)\right) \left(\tilde{x}_{\downarrow\downarrow} + \tilde{x}_{\uparrow\uparrow}\right)\right] +\frac{\eta^2_t}{4}(\tilde{x}_{\downarrow\uparrow} + \tilde{x}_{\uparrow\downarrow}) \right\rbrace . \label{maintext-probsuccess-BKprotocol}
\end{align}

The output state of the memories at the end of the BK protocol reads:
\begin{align} \label{maintext-finalstate-BKprotocol}
    &\rho_{\rm output} =  \left\lbrace p_d^2 (1-\eta_t)^2 (\tilde{x}_{\uparrow\uparrow}\ketbra{\downarrow\downarrow}{\downarrow\downarrow} + \tilde{x}_{\downarrow\uparrow}\ketbra{\uparrow\downarrow}{\uparrow\downarrow} + \tilde{x}_{\uparrow\downarrow}\ketbra{\downarrow\uparrow}{\downarrow\uparrow} +\tilde{x}_{\downarrow\downarrow}\ketbra{\uparrow\uparrow}{\uparrow\uparrow}) \right. \nonumber\\
    &\left.\quad + p_d \eta_t \left[(1-\eta_t) \left(\tilde{x}_{\downarrow\uparrow} \ketbra{\uparrow\downarrow}{\uparrow\downarrow} + \tilde{x}_{\uparrow\downarrow} \ketbra{\downarrow\uparrow}{\downarrow\uparrow} \right) + \left(1-\eta_t\left(\frac{3}{4}-\frac{\cos^2\theta}{4}\right)\right) \left(\tilde{x}_{\downarrow\downarrow} \ketbra{\uparrow\uparrow}{\uparrow\uparrow} + \tilde{x}_{\uparrow\uparrow}\ketbra{\downarrow\downarrow}{\downarrow\downarrow} \right)\right] \right. \nonumber\\
    &\left.\quad  +\frac{\eta^2_t}{2} \left[\frac{\tilde{x}_{\downarrow\uparrow}}{2}\ketbra{\uparrow\downarrow}{\uparrow\downarrow} + \frac{\tilde{x}_{\uparrow\downarrow}}{2} \ketbra{\downarrow\uparrow}{\downarrow\uparrow} +  \frac{\cos^2 \theta}{2} \left(v^*\ketbra{\uparrow\downarrow}{\downarrow\uparrow} + v \ketbra{\downarrow\uparrow}{\uparrow\downarrow}\right)\right] \right\rbrace (1-p_d)^2 \Big/ \Pr(\textrm{success}) ,
\end{align}
and its fidelity with the ideal output state, i.e. the Bell state $\ket{\Psi^+}$, is given by:
\begin{align} \label{maintext-fidelity-BKprotocol}
    \mathcal{F} 
    &=\left\lbrace \frac{\tilde{x}_{\downarrow\uparrow} + \tilde{x}_{\uparrow\downarrow}}{2} \left[p_d(1-\eta_t) + \frac{\eta_t}{2}\right]^2 + \frac{\eta_t^2}{4} \Re(v) \cos^2 \theta \right\rbrace (1-p_d)^2 \Big/ \Pr(\textrm{success}),
\end{align}
where $\Re(v)$ extracts the real part of $v$. The expressions of the parameters $\tilde{x}_{\uparrow\uparrow}$,$\tilde{x}_{\uparrow\downarrow}$, $\tilde{x}_{\downarrow\uparrow}$, $\tilde{x}_{\downarrow\downarrow}$, and $v$ are given as follows:
\begin{align}
    \tilde{x}_{\uparrow\uparrow} &= \left[1 + \sin (2 \alpha ) \cos (\alpha') \right] \left[1 + \sin (2 \beta ) \cos (\beta')\right]  \\
    \tilde{x}_{\uparrow\downarrow} &= \left[1 + \sin (2 \alpha ) \cos (\alpha') \right] \left[1 - \sin (2 \beta ) \cos (\beta')\right]   \\
    \tilde{x}_{\downarrow\uparrow} &=  \left[1 - \sin (2 \alpha ) \cos (\alpha') \right] \left[1 + \sin (2 \beta ) \cos (\beta')\right]    \\
    \tilde{x}_{\downarrow\downarrow} &=  \left[1 - \sin (2 \alpha ) \cos (\alpha') \right] \left[1 - \sin (2 \beta ) \cos (\beta')\right]  \\
    v&= \left[\cos (2 \alpha) + i\sin (2 \alpha )\sin (\alpha')\right] \left[\cos (2 \beta )-i \sin (2 \beta ) \sin (\beta')\right].
\end{align}

Notably, despite the BK protocol is based on single-photon interference events in the central node, the output state and the success rate of the protocol are independent of relative phases between the light modes from the two memories (the parameter $\varphi$ in Subsection~\ref{sec:noise_sources}). This counter-intuitive fact, also noted in \cite{exp_3m}, is due to the two-step nature of the BK protocol, which cancels out residual phase noise from the first Bell state measurement thanks to the bit flip and the second Bell state measurement.

We also separately provide the probability that the first Bell state measurement is successful (i.e., that the BK protocol does not abort in step 3.):
\begin{align}
    P_1 = \frac{1-p_d}{2} \left[p_d \norm{m_1}^2 + \frac{\eta_t}{2} (\norm{m_2}^2 + \norm{m_3}^2) + \eta_t^2 \norm{m_4}^2 \left(\frac{\cos^2 \theta + 1}{4}\right)\right],   \label{maintext-P_1}
\end{align}
where:
\begin{align} 
    \norm{m_1}^2 &=\left[\eta_t \sin (2 \alpha ) \cos (\alpha')+2-\eta_t\right] \left[\eta_t \sin (2 \beta ) \cos (\beta')+2-\eta_t\right]  \\
    \norm{m_2}^2 &= \left[1-\sin (2 \beta ) \cos (\beta')\right] \left[ \eta_t \sin (2\alpha ) \cos (\alpha')+ 2 -\eta_t \right]   \\
    \norm{m_3}^2 &= \left[1-\sin (2 \alpha ) \cos (\alpha') \right] \left[ \eta_t \sin (2\beta ) \cos (\beta')+2 - \eta_t\right]   \\
    \norm{m_4}^2 &= \left[1-\sin (2 \alpha ) \cos (\alpha') \right] \left[1-\sin (2 \beta ) \cos (\beta')\right] ,  
\end{align}
and the probability that the second Bell state measurement is successful (i.e., that the protocol does not abort in step 5.), conditioned on succeeding in the first Bell state measurement:
\begin{align}
    P_2 = \Pr(\textrm{success}) / P_1. \label{maintext-P_2}
\end{align}

\subsection{Protocol simulation with only photon losses} \label{sec:simplified_simul}

Here, we illustrate the computations required to obtain the analytical expressions of the previous Subsection. For pedagogical reasons, we report the computations in the simplified scenario where photon losses are the only error source. In other words, we set $\theta=\varphi=\alpha=\beta=p_d=0$. The general calculation including all error sources from  Subsection~\ref{sec:noise_sources} is presented in Appendix~\ref{app:calculations}.

To simplify the calculations, we employ the following results from quantum optics \cite{common-loss}. As shown in Figure \ref{fig:BK-setup}, we model the detectors inefficiency as a beam splitter with transmittance $\eta_d$ placed before each detector. The statistics and post-selected states would be unchanged if we instead place the same beam splitter (with transmittance $\eta_d$) in each of the two input arms of the balanced beam splitter. Therefore, in each input arm of the balanced beam splitter we now have a sequence of three beam splitters (ordered from closest to furthest from the beam splitter): one with transmittance $\eta_d$ modeling the detection efficiency, one with transmittance $\eta$ modeling the lossy channel, and one with transmittance $\eta_m$ modeling the photon emission efficiency of each memory. Now, we use the fact that such a sequence of beam splitters can be equivalently described by a single beam splitter, $U_{loss}(\eta_t)$, with transmittance
\begin{align}
    \eta_t =\eta_m \eta \eta_d. \label{eta_t}
\end{align}

Given this background, it is now possible to derive the success probability and fidelity of the output state of the BK protocol. Let $M_A$, $A$, and $L_A$ ($M_B$, $B$, and $L_B$) be the labels of the qubit system in the left (right) memory, the light mode in the left (right) channel, and the light mode lost in the left (right) channel, respectively.

The state of the two memories after step 2.~of the BK protocol reads:
\begin{align}
    \ket{\psi_2} = \frac{(\ket{\uparrow} + \ket{e})_{M_A}}{\sqrt{2}} \otimes  \frac{(\ket{\uparrow} + \ket{e})_{M_B}}{\sqrt{2}}.
\end{align}
In step 3.~the memories relax, releasing a photon in modes $A$ and $B$. The global state reads:
\begin{align}
    \ket{\psi_3'} =&\frac{(\ket{\uparrow}_{M_A}\ket{0}_A + \ket{\downarrow}_{M_A}\ket{1}_A)}{\sqrt{2}} \nonumber\\
    &\otimes  \frac{(\ket{\uparrow}_{M_B}\ket{0}_B + \ket{\downarrow}_{M_B}\ket{1}_B)}{\sqrt{2}}.
\end{align}
We now apply $U_{loss}(\eta_t)$ on modes $A$ and $B$. We obtain:

\begin{align} \label{eq: starting_wf}
    \ket{\psi_3''} &=\frac{1}{\sqrt{2}}\left[\ket{\uparrow}_{M_A}\ket{00}_{AL_A} + \ket{\downarrow}_{M_A}(\sqrt{\eta_t}\ket{10}_{AL_A} + \sqrt{1-\eta_t}\ket{01}_{AL_A})\right] \nonumber\\
    &\quad\otimes  \frac{1}{\sqrt{2}}\left[\ket{\uparrow}_{M_B}\ket{00}_{BL_B} + \ket{\downarrow}_{M_B}(\sqrt{\eta_t}\ket{10}_{BL_B} + \sqrt{1-\eta_t}\ket{01}_{BL_B})\right] \nonumber\\
    &= \frac{1}{2}\left[\ket{\phi}_{M_A L_A}\ket{0}_A + \sqrt{\eta_t} \ket{\downarrow}_{M_A}\ket{0}_{L_A}\ket{1}_{A} \right] \otimes \left[\ket{\phi}_{M_B L_B}\ket{0}_B + \sqrt{\eta_t} \ket{\downarrow}_{M_B}\ket{0}_{L_B}\ket{1}_{B} \right],
\end{align}
where we defined the states:
\begin{align}
   \ket{\phi}_{M_{A(B)} L_{A(B)}} = \ket{\uparrow}_{M_{A(B)}}\ket{0}_{L_{A(B)}} + \sqrt{1-\eta_t} \ket{\downarrow}_{M_{A(B)}}\ket{1}_{L_{A(B)}}. \label{phi}
\end{align}
The photons in modes $A$ and $B$ are then combined in a balanced beam splitter with output modes labeled $C$ and $D$. This transforms the state $\ket{\psi_3''}$ into:
\begin{align} \label{eq: psi3_final}
    \ket{\psi_3'''} &= \frac{1}{2}\left[\ket{\phi}_{M_A L_A}\ket{\phi}_{M_B L_B}\ket{00}_{CD} \right.\nonumber\\
    &\quad\left. + \sqrt{\eta_t} \left(\ket{\phi}_{M_A L_A} \ket{\downarrow 0}_{M_B L_B} \frac{\ket{10}_{CD}-\ket{01}_{CD}}{\sqrt{2}} + \ket{\downarrow 0}_{M_A L_A} \ket{\phi}_{M_B L_B} \frac{\ket{10}_{CD}+\ket{01}_{CD}}{\sqrt{2}}\right) \right. \nonumber\\
    &\quad\left. + \eta_t \ket{\downarrow 0}_{M_A L_A} \ket{\downarrow 0}_{M_B L_B} \frac{\ket{20}_{CD}-\ket{02}_{CD}}{\sqrt{2}}\right] \nonumber\\
    &= \frac{1}{2}\left[\ket{\phi}_{M_A L_A}\ket{\phi}_{M_B L_B}\ket{00}_{CD} \right.\nonumber\\
    &\quad\left. + \sqrt{\eta_t} \left(\frac{\ket{\phi}_{M_A L_A} \ket{\downarrow 0}_{M_B L_B} +\ket{\downarrow 0}_{M_A L_A} \ket{\phi}_{M_B L_B}}{\sqrt{2}} \ket{10}_{CD} \right.\right. \nonumber\\
    &\quad\left.\left. + \frac{-\ket{\phi}_{M_A L_A} \ket{\downarrow 0}_{M_B L_B} +\ket{\downarrow 0}_{M_A L_A} \ket{\phi}_{M_B L_B}}{\sqrt{2}}  \ket{01}_{CD}\right) \right. \nonumber\\
    &\quad\left. + \eta_t \ket{\downarrow 0}_{M_A L_A} \ket{\downarrow 0}_{M_B L_B} \frac{\ket{20}_{CD}-\ket{02}_{CD}}{\sqrt{2}}\right].
\end{align}

We now compute the probability that the left detector (collecting mode $C$) clicks. Let $\one-\ketbra{0}{0}$ be the projector on the Fock subspace with at least one photon. Then the probability that only the detector of mode $C$ clicks is given by:
\begin{align} 
    \Pr(C\checkmark) &= \Tr\left[(\one-\ketbra{0}{0})_C \otimes \ketbra{0}{0}_D  \ketbra{\psi_3'''}{\psi_3'''}\right] \nonumber\\
    &= \frac{\eta_t}{4} \norm{\frac{\ket{\phi}_{M_A L_A} \ket{\downarrow 0}_{M_B L_B} +\ket{\downarrow 0}_{M_A L_A} \ket{\phi}_{M_B L_B}}{\sqrt{2}}}^2 + \frac{(\eta_t)^2}{8} \nonumber\\
    &= \frac{\eta_t}{4} \braket{\phi|\phi}
    + \frac{(\eta_t)^2}{8} \nonumber\\
    &=  \frac{\eta_t}{4} (2-\eta_t) + \frac{(\eta_t)^2}{8} \nonumber\\
    &= \frac{\eta_t}{2} \left(1 - \frac{\eta_t}{4}\right). \label{prob-C-click}
\end{align}
In order to derive the post-selected state of the two quantum memories, when only the detector of mode $C$ clicks, we first simplify the following expression through \eqref{phi}:
\begin{align}
    \frac{\ket{\phi}_{M_A L_A} \ket{\downarrow 0}_{M_B L_B} +\ket{\downarrow 0}_{M_A L_A} \ket{\phi}_{M_B L_B}}{\sqrt{2}} = \ket{\Psi^+}_{M_A M_B} \ket{00}_{L_A L_B} + \sqrt{1-\eta_t} \ket{\downarrow\downarrow}_{M_A M_B} \ket{\Psi^+}_{L_A L_B}.
\end{align}
Then, the final state of step 3.~of the BK protocol reads as follows when conditioned on detector in mode $C$ clicking:
\begin{align}
    \rho_{M_A M_B|C\checkmark} &= \frac{1}{\Pr(C\checkmark)} \Tr_{L_A L_B C D}\left[(\one-\ketbra{0}{0})_C \otimes \ketbra{0}{0}_D  \ketbra{\psi_3'''}{\psi_3'''}\right] \nonumber\\
    &=\frac{1}{\Pr(C\checkmark)}\left[\frac{\eta_t}{4} \left(\ketbra{\Psi^+}{\Psi^+} + (1-\eta_t) \ketbra{\downarrow\downarrow}{\downarrow\downarrow} \right) + \frac{(\eta_t)^2}{8} \ketbra{\downarrow\downarrow}{\downarrow\downarrow}\right] \nonumber\\
    &= \frac{1}{\Pr(C\checkmark)}\frac{\eta_t}{2} \left[\frac{1}{2}\ketbra{\Psi^+}{\Psi^+} + \left(\frac{1}{2}-\frac{\eta_t}{4}\right) \ketbra{\downarrow\downarrow}{\downarrow\downarrow} \right] \nonumber\\
    &= \frac{1}{1-\eta_t/4} \left[\frac{1}{2}\ketbra{\Psi^+}{\Psi^+} + \left(\frac{1}{2}-\frac{\eta_t}{4}\right) \ketbra{\downarrow\downarrow}{\downarrow\downarrow} \right]. \label{rho|C-click}
\end{align}
As one can observe from the last expression, the conditional state of the memories does contain spurious contributions $\ket{\downarrow\downarrow}$ apart from the Bell state $\ket{\Psi^+}$. We remark that these are in part due to the inefficiencies of the setup, but would also be present under ideal conditions ($\eta_m =\eta= \eta_d=1$) due to the events where both photons reach the interferometer and exit the beam splitter from the same port (Hong-Ou-Mandel interference). We will see that after the second cycle of excitation and relaxation of the memories, the spurious contributions will disappear.

In a similar manner, one can compute the probability that only the detector of mode $D$ clicks:
\begin{align}
    \Pr(D\checkmark) &= \Tr\left[\ketbra{0}{0}_C \otimes (\one-\ketbra{0}{0})_D  \ketbra{\psi_3'''}{\psi_3'''}\right] \nonumber\\
    &=\frac{\eta_t}{2} \left(1 - \frac{\eta_t}{4}\right) \nonumber\\
    &= \Pr(C\checkmark), \label{prob-D-click}
\end{align}
and observe that is equal for both detectors. Similarly, the post-selected state of the two quantum memories, conditioned on detector $D$ clicking, reads:
\begin{align}
    \rho_{M_A M_B|D\checkmark} &= \frac{1}{1-\eta_t/4} \left[\frac{1}{2}\ketbra{\Psi^-}{\Psi^-} + \left(\frac{1}{2}-\frac{\eta_t}{4}\right) \ketbra{\downarrow\downarrow}{\downarrow\downarrow} \right], \label{rho|D-click}
\end{align}
which differs from \eqref{rho|C-click} only in the Bell state component, which is now $\ket{\Psi^-}=(\ket{\uparrow\downarrow} -\ket{\downarrow\uparrow})/\sqrt{2}$.

Now, we proceed with step 4.~of the BK protocol and apply the Pauli $X$ gate on the quantum memories. This induces the transitions $\ket{\uparrow} \to\ket{\downarrow}$ and $\ket{\downarrow} \to\ket{\uparrow}$. The resulting states of the quantum memories read:
\begin{align}
    \rho_{M_A M_B|\nicefrac{C}{D}\checkmark} &= \frac{1}{1-\eta_t/4} \left[\frac{1}{2}\ketbra{\Psi^{\pm}}{\Psi^{\pm}} + \left(\frac{1}{2}-\frac{\eta_t}{4}\right) \ketbra{\uparrow\uparrow}{\uparrow\uparrow} \right], \label{rho|CD-click-Xgate}
\end{align}
where we used the fact that pure states are defined modulo a global phase. 

Step 5.~in the BK protocol consists in exciting the memories with the $\pi$-pulse. This operation does not affect the spurious term $\ketbra{\uparrow\uparrow}{\uparrow\uparrow}$, which by definition is not excited. Therefore, we can focus our attention on the Bell state component $\ket{\Psi^{\pm}}$, which transforms as follows after a cycle of excitation and relaxation:
\begin{align}
    \ket{\tilde{\Psi}^{\pm}} &= \frac{1}{\sqrt{2}}\left[\ket{\uparrow\downarrow}_{M_A M_B} \ket{00}_{AL_A} (\sqrt{\eta_t} \ket{10}_{B L_B} + \sqrt{1-\eta_t} \ket{01}_{B L_B}) \right.\nonumber\\
    &\quad\left.\pm \ket{\downarrow\uparrow}_{M_A M_B} (\sqrt{\eta_t} \ket{10}_{A L_A} + \sqrt{1-\eta_t} \ket{01}_{A L_A}) \ket{00}_{B L_B}\right] \nonumber\\
    &= \frac{1}{\sqrt{2}}\left[\sqrt{1-\eta_t} \ket{00}_{AB}(\ket{01}_{L_A L_B} \ket{\uparrow\downarrow}_{M_A M_B} \pm \ket{10}_{L_A L_B} \ket{\downarrow\uparrow}_{M_A M_B}) \right.\nonumber\\
    &\quad\left. +\sqrt{\eta_t} \ket{00}_{L_A L_B} (\ket{\uparrow\downarrow}_{M_A M_B}\ket{01}_{AB} \pm \ket{\downarrow\uparrow}_{M_A M_B}\ket{10}_{AB}) \right].
\end{align}
We remark that the last expression describes the global matter-photon state before the photons are combined in the balanced beam splitter. After interfering the photons in modes $A$ and $B$ in the balanced beam splitter, the Bell state component becomes:
\begin{align}
    \ket{\dbtilde{\Psi}^{\pm}} &= \frac{1}{\sqrt{2}}\left[\sqrt{1-\eta_t} \ket{00}_{CD}(\ket{01}_{L_A L_B} \ket{\uparrow\downarrow}_{M_A M_B} \pm \ket{10}_{L_A L_B} \ket{\downarrow\uparrow}_{M_A M_B}) \right.\nonumber\\
    &\quad\left. +\sqrt{\eta_t} \ket{00}_{L_A L_B} (\ket{10}_{CD}\ket{\Psi^\pm}_{M_A M_B} -\ket{01}_{CD} \ket{\Psi^\mp}_{M_A M_B}) \right]. \label{psitildetilde}
\end{align}
From the last expression, we deduce that a click in the detector $C$ projects the memories' state onto $\ket{\Psi^\pm}$, which implies that if detector $C$ clicked in both cycles then the post-selected state is $\ket{\Psi^+}$; vice-versa, if the detector $D$ clicked in the first cycle and detector $C$ in the second, the post-selected state is $\ket{\Psi^-}$. An analogous pattern holds when the detector $D$ clicks in the second cycle. We conclude that when the same detectors (different detectors) click in both cycles, the state is $\ket{\Psi^+}$ ($\ket{\Psi^-}$).

The matter-photon state at the end of step 5.~of the BK protocol reads:
\begin{align}
    \rho'_{M_A M_B L_A L_B CD|\nicefrac{C}{D}\checkmark} &= \frac{1}{1-\eta_t/4} \left[\frac{1}{2}\ketbra{\dbtilde{\Psi}^{\pm}}{\dbtilde{\Psi}^{\pm}} + \left(\frac{1}{2}-\frac{\eta_t}{4}\right) \ketbra{\uparrow\uparrow}{\uparrow\uparrow} \otimes \ketbra{0000}{0000}_{L_A L_B CD} \right]. \label{rhoprime|CD-click}
\end{align}
Now, by post-selecting again the state on a click in detector $C$, the memories are left in the following unnormalized state when the same detector clicks in both cycles of excitation of and relaxation:
\begin{align} \label{eq: unnorm}
    \Pr(C\checkmark|C\checkmark) \rho_{M_A M_B|C\checkmark, C\checkmark} &= \Tr_{L_A L_B CD}\left[ (\one-\ketbra{0}{0})_C \otimes \ketbra{0}{0}_D \rho'_{M_A M_B L_A L_B CD|C\checkmark}\right] \nonumber\\
    &=\frac{1}{1-\eta_t/4} \left[\frac{\eta_t}{4} \ketbra{\Psi^+}{\Psi^+} \right],
\end{align}
which implies that the normalized state of the memories is the pure state $\ket{\Psi^+}$ and that the probability of detector $C$ clicking, given that it clicked in the first cycle, is:
\begin{align}
    \Pr(C\checkmark|C\checkmark) = \frac{\eta_t}{4-\eta_t}.
\end{align}
One can easily verify that:
\begin{align}
    \Pr(D\checkmark|C\checkmark) =\Pr(D\checkmark|D\checkmark) = \Pr(C\checkmark|D\checkmark) = \Pr(C\checkmark|C\checkmark)= \frac{\eta_t}{4-\eta_t}.
\end{align}
In conclusion, the state of the memories at the end of the BK protocol is $\ket{\Psi^+}$ ($\ket{\Psi^-}$) when equal (different) detectors clicked in the two excitation cycles. The last step of the protocol, when the $Z$ gate is applied, ensures that the memories are left in the state $\ket{\Psi^+}$. 

Therefore, the fidelity of the output state of the BK protocol with the Bell state $\ket{\Psi^+}$ reads:
\begin{align}
    \mathcal{F} =1,
\end{align}
which is indeed recovered from \eqref{maintext-fidelity-BKprotocol} when setting $\theta=\varphi=\alpha=\beta=p_d=0$.

The total success probability of the BK protocol is given by:
\begin{align} \label{eq: final}
    \Pr(\textrm{success}) &= \Pr(C\checkmark|C\checkmark)\Pr(C\checkmark) + \Pr(C\checkmark|D\checkmark)\Pr(D\checkmark) + \Pr(D\checkmark|C\checkmark)\Pr(C\checkmark) + \Pr(D\checkmark|D\checkmark)\Pr(D\checkmark) \nonumber\\
    &=\frac{\eta_t^2}{2},
\end{align}
which again is recovered from \eqref{maintext-probsuccess-BKprotocol} when setting $\theta=\varphi=\alpha=\beta=p_d=0$.

\section{Simulations on upgraded SeQUeNCe} \label{sec: simul_seq}

In this Subsection we integrate the analytical results for the execution of a noisy BK protocol, from Subsection~\ref{sec:analytical_model}, in our upgraded version of the quantum network simulator SeQUeNCe. Moreover, we perform additional changes to improve the performance of the simulator (Subsection \ref{sec: framework}). The resulting library is then tested and compared with the original version of SeQUeNCe (Subsection \ref{sec: comp}). Then, in Subsection \ref{sec: simul}, we verify the effects of the error parameters introduced in the code on entanglement time and fidelity. Lastly, in Subsection~\ref{sec: exp} we employ our upgraded library to replicate the results of a real experiment.

\subsection{Computational framework} \label{sec: framework}

\begin{figure}
    \centering
    \includegraphics[width=0.8\linewidth]{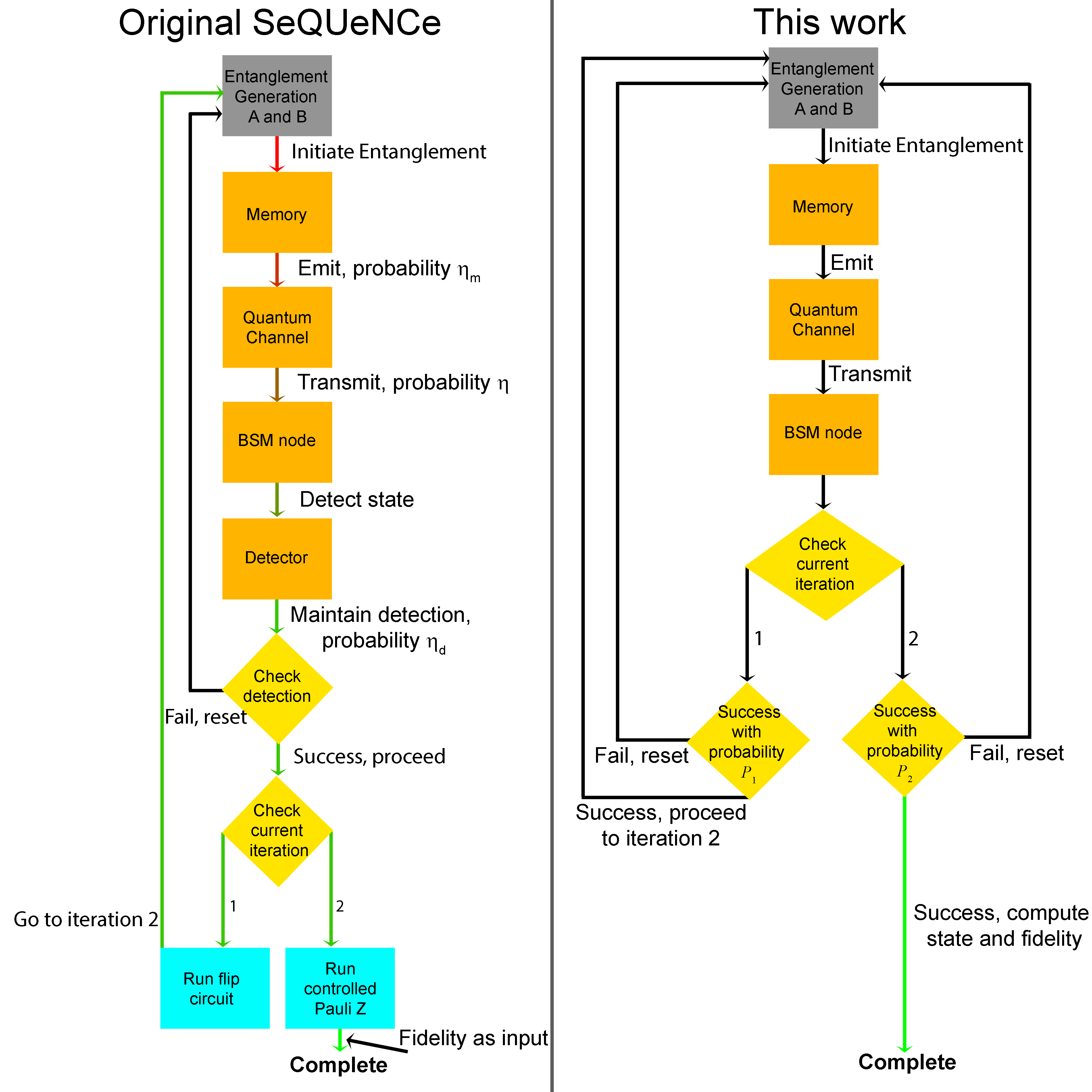}
    \caption{A simplified overview of the computational pipeline for the original SeQUeNCe (on the left) and our work (on the right). The grey and orange squares represent modules, with the orange ones being modules that represent physical components of the setup. The yellow diamonds represent if/else statements. The blue squares represent operations that require simulating a quantum circuit. The arrows represent operations. The arrow is colored if a state is being tracked, and changes color whenever an operation might change the matter-photon state. At a glance, the main difference from the original pipeline is that we do not require tracking and updating a quantum state throughout the process (black arrows on the right). Instead, the output state and probability of success of the BK protocol are computed via the analytical formulas in Subsection~\ref{sec:analytical_model}. Furthermore, we do not need to simulate a quantum circuit (lack of blue squares).\cor{This is the base pipeline, but due to the modular nature of SeQUeNCe, modifying it in order to fit a specific system should be straightforward. For instance, it would be simple to compute the main intermediate state of the process. That is, the one after the first entanglement round (Equation \eqref{eq: rho|C-step3-imperfectstateprep}), if required for intermediate operations.}}
    \label{fig:pipeline}
\end{figure}

The main idea is to employ the formulas presented in Subsection~\ref{sec:analytical_model} in order to obtain fidelity and success probability of the BK protocol while minimizing the amount of computations performed. The main differences between our approach and the standard one in SeQUeNCe are summarized in \textbf{Figure~\ref{fig:pipeline}}. In particular, the default version of SeQUeNCe attempts to replicate as much as possible the physical steps of the protocol, by keeping track of the changes to the matter-photon state caused by the three loss sources (memory efficiency, detector efficiency and channel transmittance). This is achieved through a quantum manager module which handles the state. Furthermore, the standard version of SeQUeNCe takes the fidelity of the output state of the BK protocol as an input parameter.

However, since the scope of the BK protocol within SeQUeNCe is only to provide the resulting joint state of the memories, its fidelity and a correct simulation of the time taken to perform it, there is no apparent reason to track all state changes occurring during the protocol. In fact, this would just significantly increase the computation time.

Therefore, our version of SeQUeNCe maintains the original structure only insofar as the simulation of the time required by each step, but never actually updates the state. Instead, at the end of each iteration of excitation and measurement, it utilizes the probabilities $P_1$, in Equation \eqref{maintext-P_1}, and $P_2$, in Equation \eqref{maintext-P_2}, in order to decide whether the Bell state measurement succeeded or failed. For instance, at the end of the first iteration, with probability $P_1$ the protocol proceeds to the second iteration, otherwise it registers a failure and begins a new instance of the BK protocol. The same thing happens at the end of the second iteration. When both iterations are successful, the final state and its amount of entanglement can be computed from the physical characteristics of the system by utilizing Equations \eqref{maintext-finalstate-BKprotocol} and \eqref{maintext-fidelity-BKprotocol}. This is not only faster, but seamlessly integrates the additional error sources and allows the simulator to obtain the fidelity of the output state starting from the given noise parameters, instead of taking the final state fidelity as input like in native SeQUeNCe.

An important remark regards the handling of imperfect state preparation. The most likely experimental scenario is for the setup to attempt to produce exactly the desired state in each memory initialization, while any non-ideality can be seen as the result of the stochastic nature of the state preparation. Therefore, instead of setting a fixed value for $\alpha$, $\beta$, $\alpha'$ and $\beta'$, we sample them from a normal distribution centered in $0$ (perfect state preparation) with standard deviation $0.05\pi$, unless specified otherwise. \cor{This value for the standard deviation is just a general assumption, and in real use cases should be obtained from the experimental apparatus.} On the other hand, we assume that the variables $\theta$ and $\varphi$ parameterizing the imperfect mode matching are roughly constant over many runs of the BK protocol and are hence not sampled.

In order to obtain a meaningful value for the time required for generating an entangled pair from the simulator, we need to set the nodes' distance. Indeed, the distance between the memories influences the transmission time of the photons and classical signals exchanged between the memories. In this work we set it for simplicity to $3$ meters for all simulations. Moreover, we set the decoherence time to the constant value of $10$ ms, which is much greater than $5.6\times10^{-2}$ ms, i.e., the time required for a full execution of the BK protocol, and thus does not affect the simulation results.

\subsection{Comparison with original SeQUeNCe} \label{sec: comp}
\begin{figure*}[htbp]
    \centering
    \begin{subfigure}[b]{0.31\columnwidth}
        \includegraphics[width=\linewidth]{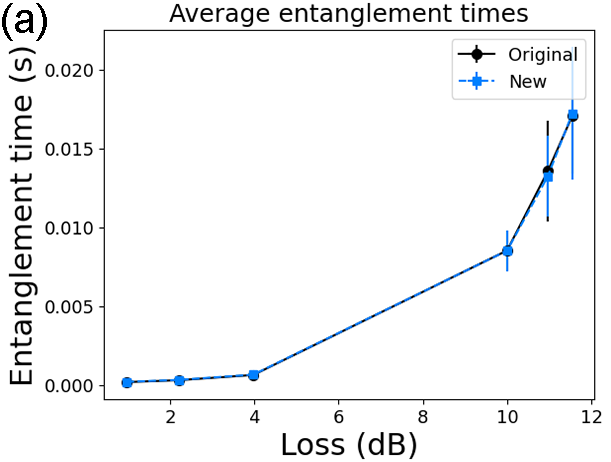}
    \end{subfigure}
    \begin{subfigure}[b]{0.31\columnwidth}
        \includegraphics[width=\linewidth]{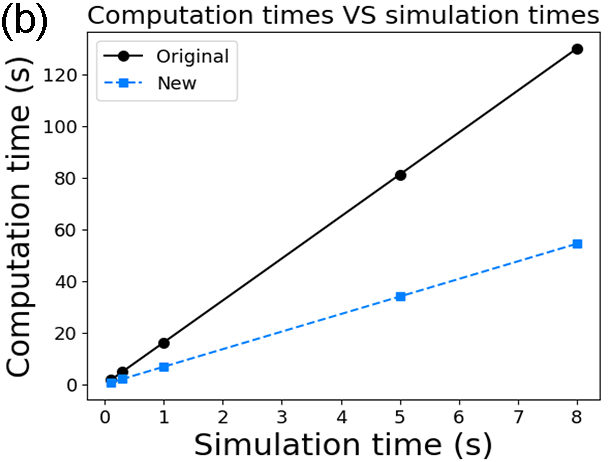}
    \end{subfigure}
    \begin{subfigure}[b]{0.31\columnwidth}
        \includegraphics[width=\linewidth]{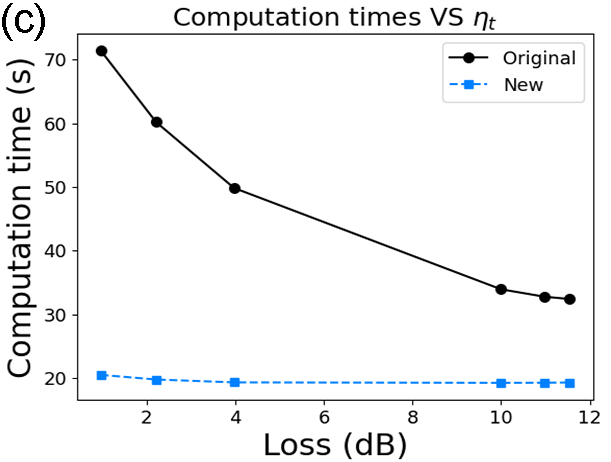}
    \end{subfigure}
    \caption{(a) Average time to obtain a successful entanglement event between two memories, as a function of the total photon loss $L = -10\log_{10}\eta_t$, where $\eta_t=\eta_m \eta \eta_d$. (b) Average computation time as a function of the total simulated time. (c) Average computation time as a function of total loss, $L$. In all plots the black solid lines are the results for the native SeQUeNCe, while the blue dashed lines are the results for our version. The computation times reported in this figure are obtained on the following CPU: AMD Ryzen Threadripper 3990X 64-Core Processor.}
    \label{fig: plots_compare}
\end{figure*}

We benchmark our code with the default version of SeQUeNCe in order to verify its correctness and its advantages over the standard library. Specifically, we run the entanglement generation protocol --i.e., the software implementation of the BK protocol-- in both versions of SeQUeNCe having fixed the total simulation time. Note that since native SeQUeNCe does not take into account additional error parameters on top of photon loss (represented by $\eta_t$), we fix the other error parameters to their ideal values in our version of the simulator. Each time entanglement generation is successful, we store the total number of attempts before entanglement was established and the elapsed time (i.e. the time required to obtain a successful entanglement event). Then, we reset the memories in order to start another instance of entanglement generation. This procedure is repeated until the total simulation time expires. The results are summarized in \textbf{Figure \ref{fig: plots_compare}}.

In Figure \ref{fig: plots_compare}a we display the average time required to obtain a successful entanglement generation event -- note that this is the simulated time, i.e. the time that an actual experiment would take, and not the computation time. This is plotted as a function of the loss $L$, defined through the parameter $\eta_t=\eta_m \eta \eta_d$ by: $L = -10\log_{10}\eta_t$. 
We observe that when there are no additional error sources, the native version of SeQUeNCe and our version predicted the same average entanglement time, for each value of $\eta_t$. Moreover, the average entanglement time increases with the total loss $L$, as expected.

In Figures \ref{fig: plots_compare}b and \ref{fig: plots_compare}c we compare the performances of the two libraries. In particular, in Figure \ref{fig: plots_compare}b we show the effects of the total simulation time on the required computation time, for fixed error parameters. For both versions the scaling is linear, but our version takes only about $40 \%$ of the computation time required by the native version of SeQUeNCe, with a slope of $6.8$ compared to $16.2$ of the default library. Moreover, it is interesting to look at the effects of $\eta_t$ on the computation time in Figure \ref{fig: plots_compare}c, for a fixed simulation time. In general, the less losses there are the more time the simulation takes, with the effects being significantly more pronounced for the original SeQUeNCe. This is probably because a greater value of $\eta_t$ makes successful entanglement attempts more common, thus requiring more computations to generate the final state. \cor{We expect that the observed behavior will maintain its better scaling as network size increases, compounding the advantage with the amount of nodes.}

The last comparative analysis between the two versions of SeQUeNCe regards the protocol's success probability. The success probability can be inferred from the average number of entanglement generation protocol runs ($\bar{n}$) before an entangled pair is generated. We recall that the protocol can fail at the end of the first iteration with probability $1-P_1$ or at the end of the second iteration with probability $1-P_2$, while it succeeds with probability $q = P_1 P_2$. The number $n$ of protocol runs before an entanglement success, also called entanglement attempts from now on, follows a geometric distribution: $\Pr(n) = (1-q)^{n-1} q$, with expected value: $\bar{n} = 1/q$. Therefore, we obtain an estimation of the success probability of the entanglement generation protocol from the observed average number of attempts ($\bar{n}$) with the bias-corrected formula:
\begin{equation} \label{eq: p_bias_corrected}
    \hat{q} = \frac{1}{\bar{n}} - \frac{\bar{n}^{-1}\left(1-\bar{n}^{-1}\right)}{n}.
\end{equation}
We apply the above formula to estimate the success probability of the BK protocol from the data of both versions of SeQUeNCe. The results are summarized in Table \ref{tab: probabilities} and compared with the true success probability, given by: $q=\eta_t^2/2$.

\begin{table}[htp]
    \centering
    \begin{tabular}{@{}llll@{}}
    \hline
    $\eta_t$ & $\hat{q}_{\rm native}$ & $\hat{q}_{\rm upgraded}$ & $q$ \\
    \hline
    0.07  & 0.00246 & 0.00244 & 0.00245 \\

    0.08 & 0.00312 & 0.00312 & 0.0032 \\

    0.1 & 0.00503 & 0.00504 & 0.005 \\

    0.4 & 0.0826 & 0.0808 & 0.08 \\

    0.6 &  0.199 & 0.196 & 0.18 \\

    0.8 & 0.335 & 0.332 & 0.32\\
    \hline
    \end{tabular}
    \caption{Estimated success probability of entanglement generation from data taken with native SeQUeNCe and our version. We also report the true value of the success probability of the BK protocol, given by: $\eta_t^2/2$.}
    \label{tab: probabilities}
\end{table}

In general, we can see that three probabilities in Table \ref{tab: probabilities} are extremely close, with an error of at most $5 \%$. This discrepancy is mostly between the estimated probabilities and the expected value and it is likely caused by the fact that the limited simulation time is not enough to capture the tail of the geometric distribution, leading to a further bias that is not accounted for in Equation \eqref{eq: p_bias_corrected}, causing an overestimation of the success probability. Regardless, our results are strong evidence that the two simulators agree with each other and with the theory.

\subsection{Simulating noisy entanglement generation} \label{sec: simul}

\begin{figure*}[htbp]
    \centering
    \begin{subfigure}[b]{0.48\columnwidth}
        \includegraphics[width=\linewidth]{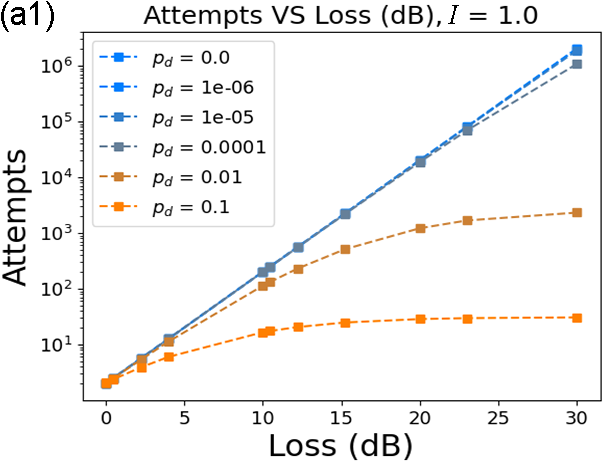}
    \end{subfigure}
    \begin{subfigure}[b]{0.48\columnwidth}
        \includegraphics[width=\linewidth]{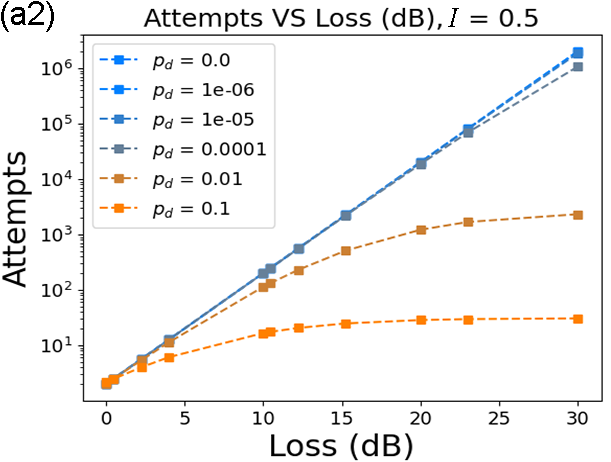}
    \end{subfigure} \\
    \begin{subfigure}[b]{0.48\columnwidth}
        \includegraphics[width=\linewidth]{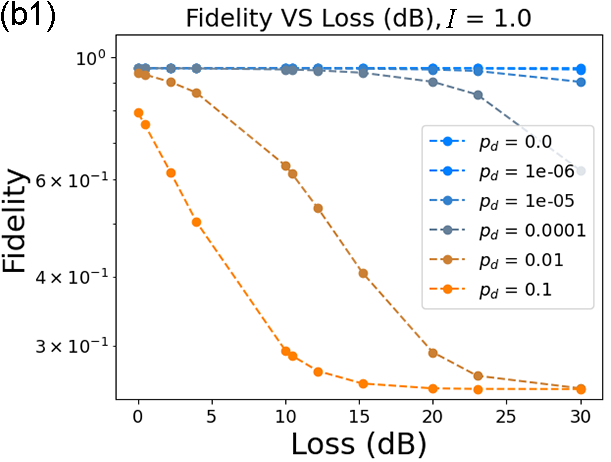}
    \end{subfigure}
    \begin{subfigure}[b]{0.48\columnwidth}
        \includegraphics[width=\linewidth]{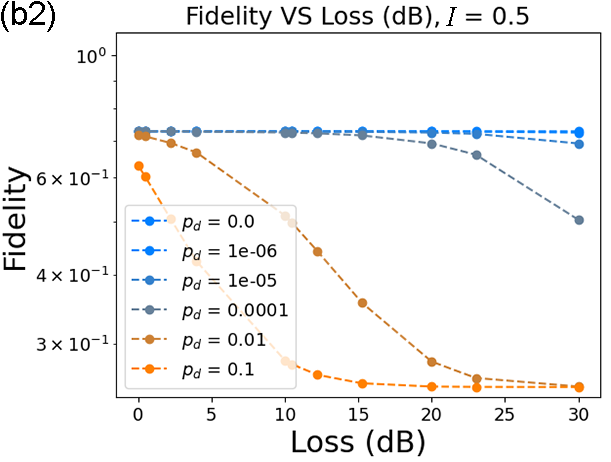}
    \end{subfigure}
    \caption{(a) Average number of entanglement attempts as a function of total loss $L$, given by: $L = -10\log_{10}\eta_t$, for (a1) indistinguishable photons ($I = 1$ and $\theta = 0$)  and (a2) photons with indistinguishability $I = 0.5$ ($\theta = \frac{\pi}{4}$) . (b) Average fidelity of entangled pairs as a function of $L$ for (b1) indistinguishable photons ($I=1$) and (b2) photons with $I = 0.5$. In all plots the results are displayed for various values of dark count probability, from $p_d = 0$ (blue) to $p_d = 0.1$ (orange)}\
    \label{fig: plots_params1}
\end{figure*}

In this Subsection we employ our upgraded quantum network simulator to investigate the impact of various error sources on the performance of the entanglement generation protocol. In particular, we focus on two key performance metrics, namely the average number of attempts (i.e., runs of the BK protocol) to successfully establish entanglement and the output state fidelity. We utilize the number of entanglement attempts as a measure of the time required for establishing entanglement, which is only affected by error parameters and not by other characteristics of the experimental setup such as the distance between memories. Firstly, in \textbf{Figure \ref{fig: plots_params1}} are shown the effects of losses ($\eta_t$) and dark counts ($p_d$) on the metrics, for two selected values of indistinguishability ($I$). In particular, the metrics are plotted as a function of the overall loss $L$, measured in dB, and given by $L = -10\log_{10}\eta_t$.

Figures \ref{fig: plots_params1}a1 and \ref{fig: plots_params1}a2 show the average number of entanglement attempts for various values of the dark count probability ($p_d$) and for two fixed values of $\theta$, which correspond to photon indistinguishability values of $I = 1.0$ and $0.5$ respectively (see Appendix \ref{sec: photon_ind} for details on the relationship between $I$ and $\theta$).

First of all, we can see at a glance that the effect of different values of $I$ on entanglement time is basically negligible when compared to the effects of photon loss and dark counts. Moreover, we observe that higher losses result in a drastic increase of the entanglement time. The dark count probability has the opposite effect, as dark counts might cause the setup to register a click when no photon is present. Hence, higher values of $p_d$ decrease the average number of entanglement attempts. 

Importantly, we observe two different regimes in the plots of Figures \ref{fig: plots_params1}a1 and \ref{fig: plots_params1}a2. For $p_d \ll \eta_t$, the average number of attempts increases exponentially with the loss $L$. While for $p_d \sim \eta_t$ and $L \geq 20$ dB, the number of attempts is almost constant in $L$ since most of the detections causing the setup to herald the generation of entanglement are dark counts.

Figures \ref{fig: plots_params1}b1 and \ref{fig: plots_params1}b2 plot the fidelity of the output state against the overall loss $L$. We deduce from the plots that losses have no effect on the fidelity in absence of dark counts. This is in agreement with the analytical formula for the fidelity, Equation \eqref{maintext-fidelity-BKprotocol}, which for $p_d=0$ becomes:

\begin{align} \label{eq: fidelity_nopd}
    \mathcal{F}_{p_d=0} &= \left\lbrace \frac{\tilde{x}_{\downarrow\uparrow} + \tilde{x}_{\uparrow\downarrow}}{2} \left[\frac{\eta_t}{2}\right]^2 + \frac{\eta_t^2}{4} \Re(v) \cos^2 \theta \right\rbrace \Big/ \Pr(\textrm{success})\\
    &=  \left\lbrace \frac{\tilde{x}_{\downarrow\uparrow} + \tilde{x}_{\uparrow\downarrow}}{2} \left[\frac{\eta_t}{2}\right]^2 + \frac{\eta_t^2}{4} \Re(v) \cos^2 \theta \right\rbrace \Big/ \left[\frac{\eta^2_t}{4}\left(\tilde{x}_{\downarrow\uparrow} + \tilde{x}_{\uparrow\downarrow}\right)\right] \\
    &= \left\lbrace \frac{\tilde{x}_{\downarrow\uparrow} + \tilde{x}_{\uparrow\downarrow}}{2} + \Re(v) \cos^2 \theta \right\rbrace \Big/ \left(\tilde{x}_{\downarrow\uparrow} + \tilde{x}_{\uparrow\downarrow}\right)
    \end{align}

and does not depend on $\eta_t$. More intuitively, this agrees with the functioning of the BK protocol discussed in Subsection \ref{sec:protocol}, where losses can either cause the protocol to abort or generate the spurious contributions in the state of the two memories, which are removed
by the second iteration of the protocol. In either case, the fidelity of the output state is not affected.

Conversely, for $p_d \gtrsim 10^{-3}$, as the losses increase it is more likely that a detection is caused by a dark count, which  significantly decreases the fidelity. In particular, for high values of dark counts the fidelity tends to  $1/4$, which is the fidelity of the maximally mixed state \cor{$\frac{\one}{4} = \nicefrac{\left(\ketbra{\Psi^+}{\Psi^+}+\ketbra{\Psi^-}{\Psi^-}+\ketbra{\Phi^+}{\Phi^+}+\ketbra{\Phi^-}{\Phi^-}\right)}{4}$, in fact the fidelity with $\ketbra{\Psi^+}{\Psi^+}$ will be $\mathcal{F} =  \braket{\Psi^+|\frac{1}{4}\one|\Psi^+} = \frac{1}{4}$}. We also observe the effect of photon distinguishability on the plots in Figure \ref{fig: plots_params1}b1 and \ref{fig: plots_params1}b2, which is that of translating all the plot lines downwards. Lastly, it is important to note that in Figure \ref{fig: plots_params1}b1, when $\eta_t=1, p_d = 0, I=1$ the fidelity does not reach $1$. That is because there are still the effects of imperfect state preparation slightly reducing it compared to the ideal value.

\begin{figure*}[htbp]
    \centering
    \begin{subfigure}[b]{0.48\columnwidth}
        \includegraphics[width=\linewidth]{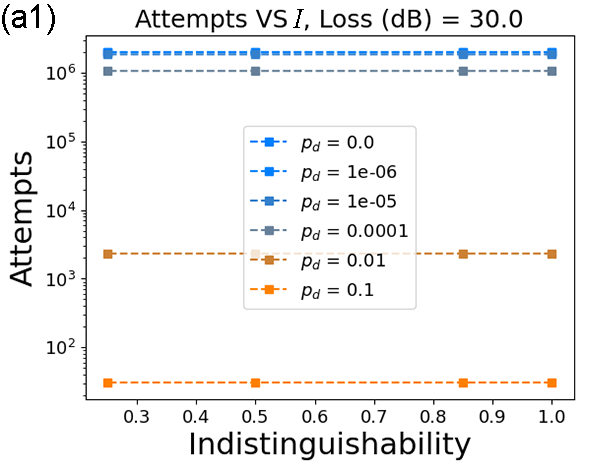}
    \end{subfigure}
    \begin{subfigure}[b]{0.48\columnwidth}
        \includegraphics[width=\linewidth]{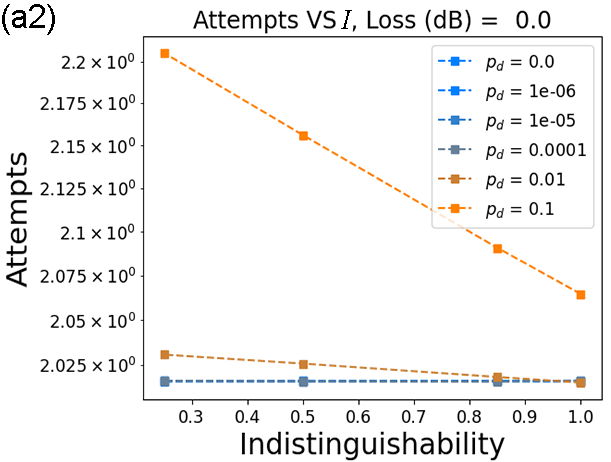}
    \end{subfigure} \\
    \begin{subfigure}[b]{0.48\columnwidth}
        \includegraphics[width=\linewidth]{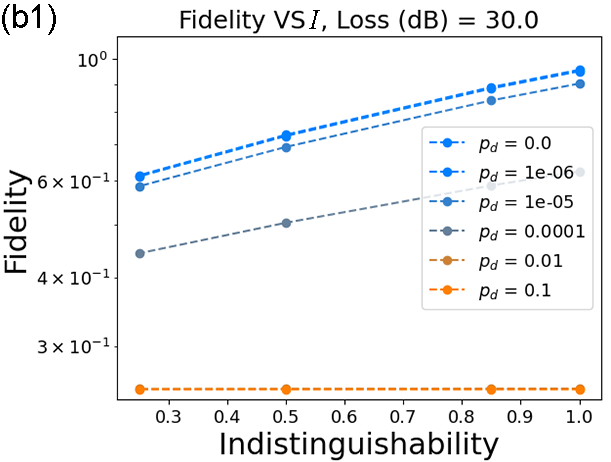}
    \end{subfigure}
    \begin{subfigure}[b]{0.48\columnwidth}
        \includegraphics[width=\linewidth]{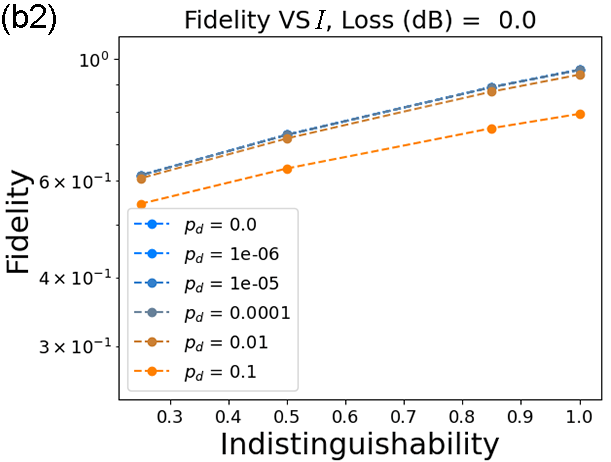}
    \end{subfigure}
    \caption{(a) Average number of entanglement attempts as a function of $I$ for (a1) $\eta_t = 10^{-3}$  and (a2) $\eta_t = 1$. (b) Average fidelity of entangled pairs as a function of $I$ for (a1) $\eta_t = 10^{-3}$  and (a2) $\eta_t = 1$. In all plots the results are displayed for various values of dark count probability, from $p_d = 0$ (blue) to $p_d = 0.1$ (orange)}\
    \label{fig: plots_params2}
\end{figure*}

We can now look more deeply at the effects of photon indistinguishability. To do so, we set $I$ as the independent variable and set two values of $\eta_t$, obtaining \textbf{Figure \ref{fig: plots_params2}}.

Figures \ref{fig: plots_params2}a1 and \ref{fig: plots_params2}a2  plot the average number of entanglement attempts as a function of the photon indistinguishability parameter $I$, for two fixed values of $\eta_t$ (high and low losses). At high losses (Figure \ref{fig: plots_params2}a1), photon indistinguishability does not have a noticeable effect on the number of attempts, which remains approximately constant with varying $I$.
Conversely, in a regime of no losses (Figure \ref{fig: plots_params2}a2) and high dark counts, there is a visible effect on the average number of  entanglement attempts, which decreases as $I \to 1$.

This can be explained by the fact that photon indistinguishability can only influence the overall success of the BK protocol (i.e., observing two consecutive single clicks in the two iterations) when both memories emit a photon in one of the iterations and the photons reach the central node. However, since this event is negligible at high losses, we observe no change in the average number of attempts for varying $I$.
Vice-versa, at very low losses, distinguishable photons from both memories can reach the central node and cause a double click, after which the protocol aborts and starts with a new attempt, thereby increasing the average number of attempts. However, in the BK protocol, even indistinguishable photons coming from both memories can cause the protocol to abort in the second iteration (the removal of the spurious contributions $\ket{\downarrow}\ket{\downarrow}$) if no dark count occurs. Hence, only in a regime of high dark counts and low loss, we expect to observe an influence of photon distinguishability on the number of attempts, whereby the latter is increased for more distinguishable photons.

Figures \ref{fig: plots_params2}b1 and \ref{fig: plots_params2}b2 plot the fidelity against photon indistinguishability and we observe a clear trend of increasing fidelity as the photons from the two memories become more indistinguishable. In this case, the influence of indistinguishability is larger than on the average number of attempts. This is because the output state fidelity is a quantity computed on the instances where the protocol succeeded, i.e., a single click occurred both in the first and second iteration. In these instances, the fidelity is affected by the photon distinguishability even in the events where only one memory emitted a photon, since distinguishable photons would provide information on which memory emitted the photon, decreasing the fidelity of the memory state with $\ket{\Psi^+}$. \cor{Furthermore, comparing the results in Figure \ref{fig: plots_params1} and \ref{fig: plots_params2}, transmittance is the main source that affects entanglement time. On the other hand, I is the dominant constraint on fidelity for low $p_d$ ($p_d \ll 10^{-4}$), while transmittance is more relevant for high $p_d$ ($p_d \gg 10^{-4}$). Of the two, the low-$p_d$ regime is the most likely in applications.}

\subsection{Replication of experimental results} \label{sec: exp}

We benchmark the capabilities of our simulator by using it to reproduce experimental results from a known experiment. In particular, we run our entanglement generation simulation when using the experimental parameters reported in  \cite{exp_3m} and compare the resulting average entanglement time and fidelity with those of the actual experiment. It is important to note that while our analytical model and upgraded simulator do capture the main error sources in a setup like that of \cite{exp_3m}, there are still some noise effects that are not being modeled.

\begin{table}[htp]
    \centering
    \begin{tabular}{@{}ll@{}}
    \hline
    Parameter & Value \\
    \hline
    $\eta_t$ & 4 $\times 10^{-4}$ \\

    $I$ & $0.80 \pm 0.05$ \\

    $p_d$ & 5.7 $\times 10^{-6}$ \\
    \hline
    \end{tabular}
    \caption{Experimental parameters taken from \cite{exp_3m}}
    \label{tab: parmeters}
\end{table}

\begin{table}[htp]
    \centering
    \begin{tabular}{@{}ll@{}}
    \hline
    Metric & Value \\
    \hline

    Average time per success & $\sim 10 $ min \\

    Average fidelity & $0.73 \pm 0.04$ \\
    \hline
    \end{tabular}
    \caption{Experimental performance metrics taken from \cite{exp_3m}}
    \label{tab: perf}
\end{table}

The parameters for the experimental setup under study are summarized in Table \ref{tab: parmeters}, while the performance metrics that we replicate are in Table \ref{tab: perf}.

In the following we explain how we extracted the parameters from Table \ref{tab: parmeters}. The parameter $\eta_t$ is reported explicitly by the paper as the overall detection probability per pulse. The background contributions are reported by the paper to be about $\eta_t/70$ and are modeled in our framework with $p_d$.  The visibility of the two-photon interference, which we also call photon indistinguishability in this work, is reported to be $0.80 \pm 0.05$. No measure of the imprecision of the state preparation is given, except for its effect on the fidelity of the output state. Therefore, we set $\alpha = \beta = \alpha' = \beta' = 0$, and then later subtract the effects of imperfect state preparation from the state fidelity.

Using the parameters in Table \ref{tab: parmeters}, our analytical model returns a value for the fidelity of $\mathcal{F}=0.83$. Then, we modify it to account for the effects of additional error contributions, present in the experiment, which are not accounted for in our model. In particular, errors in the microwave pulses reportedly decrease the fidelity by $3.5 \%$, imperfect state initialization by $2 \%$, spin flips during the optical excitation by $1\%$ and spin decoherence by less than $1\%$. Subtracting these contributions from the fidelity obtained from our model we obtain  $\mathcal{F}=0.76$, which is within one standard deviation of the experimentally reported fidelity of $0.73 \pm 0.04$. 
The other metric we evaluate is the average time needed for successfully establishing entanglement. We run the entanglement generation protocol until $50$ successes are reached, obtaining an average time per success of $8.5$ minutes. In \cite{exp_3m}, the average time is reported in the order of $10$ minutes, which is consistent with our result.

\section{Conclusion} \label{sec:conclusions}

We implemented a generalized, realistic and efficient way to simulate entanglement heralding in quantum networks, by analytically deriving the relevant figures of merit in the presence of the most common error sources and streamlining the computation pipeline in a state-of-the-art simulator of quantum networks (SeQUeNCe).

The resulting simulator is able to take into account photon losses, imperfect mode matching, detector dark counts and imperfect state preparation as potential error sources, obtaining realistic estimates of the success probability of each iteration of the protocol, the final entangled state and its fidelity with the ideal Bell state.
Furthermore, our upgraded simulator cuts down on the computation time required by $60\%$ compared to the native simulator. The upgraded simulator is benchmarked with both the native version of the simulator and with real experimental results. These benchmarks allowed us to quantify the improvement in computation time, while verifying that the results are accurate both with respect to experiments and -- when setting our additional error sources to zero -- with respect to the original library.

Our results allow for more efficient and more accurate simulations of quantum networks, enhancing scalability, and can be naturally extended to different types of entanglement protocols and experimental setups, such as single-rail and dual-rail Bell state measurements or configurations with absorbing memories. Ultimately, our software upgrades on the simulator SeQUeNCe can allow one to test, already today, high-level applications of the future quantum internet on a realistic digital testbed.


\medskip
\textbf{Acknowledgments} \par
The authors acknowledge funding from Leonardo S.p.A. in the context of the project "Quantum error correction for edge and distributed quantum computing" for the Intersectoral Doctorate for Industrial Innovation (PNRR – DM 630/2024) and from Project QMLCyber funded by Leonardo of the Consiglio Nazionale delle Ricerche (CNR)
\medskip

\clearpage

\bibliography{bibliography}

\appendix

\section{Analytical model: calculations} \label{app:calculations}

In this Appendix we present the analytical calculations leading to the expressions for the fidelity of the output state and the success probability of the BK protocol, reported in Subsection~\ref{sec:analytical_model}. In this case, we account for all noise sources from Subsection~\ref{sec:noise_sources}.

We start by modifying the state $\ket{\psi_3''}$ in Eq.~\eqref{eq: starting_wf} by accounting for imperfect mode matching (phase and polarization mismatch). The resulting state reads:
\begin{align} \label{psi3''}
    \ket{\psi_3''} &= \frac{1}{2}\left[\ket{\phi}_{M_A L_A}\ket{0}_A + e^{i\varphi}\sqrt{\eta_t} \ket{\downarrow}_{M_A}\ket{0}_{L_A}\left(\ket{1}_{A} \cos{\theta} + \ket{1}_{A_{\perp}} \sin{\theta} \right)\right] \nonumber\\
    &\quad\otimes \left[\ket{\phi}_{M_B L_B}\ket{0}_B + \sqrt{\eta_t} \ket{\downarrow}_{M_B}\ket{0}_{L_B}\ket{1}_{B} \right],
\end{align}
where the mode $A$ presents a mismatch of $\varphi$ in the phase and a mismatch of angle $\theta$ in the polarization, with respect to mode $B$. The state $\ket{\phi}$ is given in \eqref{phi}.

We proceed by considering the imperfect state preparation of the left memory \eqref{eq: imperfect_state_prepA} and the right memory \eqref{eq: imperfect_state_prepB}, which can in principle differ. To this aim, we generalize the state \eqref{psi3''} to account for the four possible sign combinations generated by the $\ket{\pm}$ contributions in each memory. We define:
\begin{align} \label{eq: psi_3_imperfect_jk}
    \ket{\psi_{3|jk}''} = &\frac{1}{2}\left[\ket{\phi^j}_{M_A L_A}\ket{0}_A + (-1)^j e^{i\varphi}\sqrt{\eta_t} \ket{\downarrow}_{M_A}\ket{0}_{L_A}\left(\ket{1}_{A} \cos{\theta} + \ket{1}_{A_{\perp}} \sin{\theta} \right)\right] \nonumber\\
    &\otimes \left[\ket{\phi^k}_{M_B L_B}\ket{0}_B + (-1)^k \sqrt{\eta_t} \ket{\downarrow}_{M_B}\ket{0}_{L_B}\ket{1}_{B} \right],
\end{align}
where we defined:
\begin{align}
   \ket{\phi^j}_{M_{A(B)} L_{A(B)}} = \ket{\uparrow}_{M_{A(B)}}\ket{0}_{L_{A(B)}} + (-1)^j \sqrt{1-\eta_t} \ket{\downarrow}_{M_{A(B)}}\ket{1}_{L_{A(B)}}. \label{phi_j}
\end{align}
We note that $\ket{\psi_{3|00}''}$ coincides with the original expression in \eqref{psi3''}. Then, by integrating \eqref{eq: imperfect_state_prepA} and \eqref{eq: imperfect_state_prepB} into the calculation, we obtain the following expression for the state $\ket{\psi_3''}$ that accounts for imperfect mode matching and imperfect state preparation:
\begin{align} \label{eq: psi_3_imperfect}
    \ket{\psi_{3}''} &= \cos\alpha\cos\beta \ket{\psi_{3|00}''} + e^{i\beta'}\cos\alpha\sin\beta \ket{\psi_{3|01}''} + e^{i\alpha'}\sin\alpha\cos\beta \ket{\psi_{3|10}''} + e^{i(\alpha' + \beta')}\sin\alpha\sin\beta \ket{\psi_{3|11}''}.
\end{align}

After modes $A$ and $B$ undergo the balanced beam splitter, the state in the last expression becomes:
\begin{align} \label{eq: psi_3_imperfect2}
    \ket{\psi_{3}'''} &= \cos\alpha\cos\beta \ket{\psi_{3|00}'''} + e^{i\beta'}\cos\alpha\sin\beta \ket{\psi_{3|01}'''} + e^{i\alpha'}\sin\alpha\cos\beta \ket{\psi_{3|10}'''} + e^{i(\alpha' + \beta')}\sin\alpha\sin\beta \ket{\psi_{3|11}'''},
\end{align}
where each component of the form $\ket{\psi_{3|jk}'''}$ is given by:
\begin{align} \label{eq: psi_3_imperfect_jk2}
    \ket{\psi_{3|jk}'''} &= \frac{1}{2}\left[\ket{\phi^j}_{M_A L_A}\ket{\phi^k}_{M_B L_B}\ket{0000}_{CD C_{\perp} D_\perp}  + (-1)^k\sqrt{\eta_t} \ket{\phi^j}_{M_A L_A} \ket{\downarrow 0}_{M_B L_B} \frac{\ket{1000}_{CD C_\perp D_{\perp}}-\ket{0100}_{CD C_\perp D_{\perp}}}{\sqrt{2}} \right.\nonumber\\
    &\quad\left.+ (-1)^je^{i\varphi}\sqrt{\eta_t} \ket{\downarrow 0}_{M_A L_A}\ket{\phi^k}_{M_B L_B} \right. \nonumber\\
    &\quad\left.\hspace{2cm}\left(\cos\theta \frac{\ket{1000}_{CD C_\perp D_{\perp}}+\ket{0100}_{CD C_\perp D_{\perp}}}{\sqrt{2}} + \sin\theta \frac{\ket{0010}_{CD C_\perp D_{\perp}}+\ket{0001}_{CD C_\perp D_{\perp}}}{\sqrt{2}}\right) \right. \nonumber\\
    &\quad\left.+(-1)^{j+k}e^{i\varphi}\eta_t \ket{\downarrow 0}_{M_A L_A} \ket{\downarrow 0}_{M_B L_B} \left(\cos\theta \frac{\ket{2000}_{CD C_\perp D_{\perp}}-\ket{0200}_{CD C_\perp D_{\perp}}}{\sqrt{2}} \right.\right. \nonumber\\
    &\quad\left.\left.+ \sin\theta \frac{\ket{1010}_{CD C_\perp D_{\perp}} + \ket{1001}_{CD C_\perp D_{\perp}} - \ket{0110}_{CD C_\perp D_{\perp}} - \ket{0101}_{CD C_\perp D_{\perp}} }{2}\right)\right].
\end{align}

The full state can be rewritten as:

\begin{align} \label{eq: psi_3_with_m}
    \ket{\psi_{3}'''} &= \frac{1}{2}\left[\ket{m_1}_{M_A L_A M_B L_B} \ket{0000}_{CD C_{\perp} D_\perp}  +\sqrt{\eta_t}\ket{m_2}_{M_A L_A M_B L_B} \frac{\ket{1000}_{CD C_\perp D_{\perp}}-\ket{0100}_{CD C_\perp D_{\perp}}}{\sqrt{2}} \right.\nonumber\\
    &\quad\left.+ e^{i\varphi}\sqrt{\eta_t}\ket{m_3}_{M_A L_A M_B L_B} \left(\cos\theta \frac{\ket{1000}_{CD C_\perp D_{\perp}}+\ket{0100}_{CD C_\perp D_{\perp}}}{\sqrt{2}} + \sin\theta \frac{\ket{0010}_{CD C_\perp D_{\perp}}+\ket{0001}_{CD C_\perp D_{\perp}}}{\sqrt{2}}\right) \right. \nonumber\\
    &\quad\left.+ e^{i\varphi}\eta_t\ket{m_4}_{M_A L_A M_B L_B} \left(\cos\theta \frac{\ket{2000}_{CD C_\perp D_{\perp}}-\ket{0200}_{CD C_\perp D_{\perp}}}{\sqrt{2}} \right.\right. \nonumber\\
    &\quad\left.\left.+ \sin\theta \frac{\ket{1010}_{CD C_\perp D_{\perp}} + \ket{1001}_{CD C_\perp D_{\perp}} - \ket{0110}_{CD C_\perp D_{\perp}} - \ket{0101}_{CD C_\perp D_{\perp}} }{2}\right)\right],
\end{align}
where we introduced the following unnormalized states:
\begin{align}
       \ket{m_1}_{M_A L_A M_B L_B} =  &\cos\alpha\cos\beta \ket{\phi^0}_{M_A L_A}\ket{\phi^0}_{M_B L_B} +  e^{i\beta'}\cos\alpha\sin\beta
       \ket{\phi^0}_{M_A L_A}\ket{\phi^1}_{M_B L_B} \nonumber\\
       &+  e^{i\alpha'}\sin\alpha\cos\beta
       \ket{\phi^1}_{M_A L_A}\ket{\phi^0}_{M_B L_B} + e^{i(\alpha' + \beta')}\sin\alpha\sin\beta
       \ket{\phi^1}_{M_A L_A}\ket{\phi^1}_{M_B L_B}\\
       \ket{m_2}_{M_A L_A M_B L_B} =  &\left(\cos\alpha\cos\beta  -  e^{i\beta'}\cos\alpha\sin\beta \right)
       \ket{\phi^0}_{M_A L_A}\ket{\downarrow 0}_{M_B L_B} \nonumber\\
       &+  \left(e^{i\alpha'}\sin\alpha\cos\beta
       - e^{i(\alpha' + \beta')}\sin\alpha\sin\beta \right)
       \ket{\phi^1}_{M_A L_A}\ket{\downarrow 0}_{M_B L_B}\\
       \ket{m_3}_{M_A L_A M_B L_B} =  &\left(\cos\alpha\cos\beta - e^{i\alpha'}\sin\alpha\cos\beta\right)
       \ket{\downarrow 0}_{M_A L_A}\ket{\phi^0}_{M_B L_B}   \nonumber\\
       &+ \left(e^{i\beta'}\cos\alpha\sin\beta
        - e^{i(\alpha' + \beta')}\sin\alpha\sin\beta\right)
       \ket{\downarrow 0}_{M_A L_A}\ket{\phi^1}_{M_B L_B}\\
       \ket{m_4}_{M_A L_A M_B L_B} =  &\left(\cos\alpha\cos\beta -  e^{i\beta'}\cos\alpha\sin\beta -  e^{i\alpha'}\sin\alpha\cos\beta
        + e^{i(\alpha' + \beta')}\sin\alpha\sin\beta\right)
       \ket{\downarrow 0}_{M_A L_A}\ket{\downarrow 0}_{M_B L_B}.
\end{align}
We observe that the polarization mismatch of the incoming modes affects the Hong-Ou-Mandel interference, such that new events where both outgoing modes contain a photon are present, compared to the scenario with no mode mismatch (c.f. Sec.~\ref{sec:simplified_simul}). This introduces double-click events, which cause the BK protocol to abort more frequently.

At this point, we compute the norm squared for each of these states in order to obtain the success probability. We obtain:
\begin{align} 
    \norm{m_1}^2 &=\left[\eta_t \sin (2 \alpha ) \cos (\alpha')+2-\eta_t\right] \left[\eta_t \sin (2 \beta ) \cos (\beta')+2-\eta_t\right] \label{eq: norm_m1} \\
    \norm{m_2}^2 &= \left[1-\sin (2 \beta ) \cos (\beta')\right] \left[ \eta_t \sin (2\alpha ) \cos (\alpha')+ 2 -\eta_t \right]  \label{eq: norm_m2} \\
    \norm{m_3}^2 &= \left[1-\sin (2 \alpha ) \cos (\alpha') \right] \left[ \eta_t \sin (2\beta ) \cos (\beta')+2 - \eta_t\right]   \label{eq: norm_m3} \\
    \norm{m_4}^2 &= \left[1-\sin (2 \alpha ) \cos (\alpha') \right] \left[1-\sin (2 \beta ) \cos (\beta')\right] .  \label{eq: norm_m4}
\end{align}

This results in the following probability of only detector $C$ (or detector $D$) clicking:
\begin{align} 
    \Pr(C\checkmark) &= \Tr\left[(\one_C \otimes \one_{C_\perp }-(1-p_d)\ketbra{00}{00}_{C C_\perp }) \otimes (1-p_d)\ketbra{00}{00}_{D D_\perp}  \ketbra{\psi_3'''}{\psi_3'''}\right] \nonumber\\
    &= \frac{1-p_d}{4} \left[p_d \norm{m_1}^2 + \frac{\eta_t}{2} \norm{\ket{m_2} + e^{i\varphi}\cos\theta \ket{m_3}}^2 + \eta_t \frac{\sin^2 \theta}{2} \norm{m_3}^2  + \eta^2_t \norm{m_4}^2 \left(\frac{\cos^2\theta}{2} + \frac{\sin^2 \theta}{4} \right) \right] \nonumber\\
    &= \frac{1-p_d}{4} \left[p_d \norm{m_1}^2 + \frac{\eta_t}{2} (\norm{m_2}^2 + \norm{m_3}^2) + \eta_t^2 \norm{m_4}^2 \left(\frac{\cos^2 \theta + 1}{4}\right)\right], \label{eq: prob_C_imperfect} \\
    \Pr(D\checkmark) &= \Pr(C\checkmark).
\end{align}

We can now compute the final state of step 3. of the BK protocol, conditioned on only detector $C$ (or $D$) clicking:
\begin{align} \label{eq: final_state_1_m}
   \rho_{M_A M_B|\nicefrac{C}{D}\checkmark} 
    &=\frac{1-p_d}{4\Pr(C\checkmark)} \nonumber\\
    &\quad\quad\left[ p_d \rho_{m_1} + \frac{\eta_t}{2} (\rho_{m_2} + \rho_{m_3}) \pm \frac{\eta_t \cos\theta}{2}   \Tr_{L_A L_B} \left[e^{-i \varphi}  \ket{m_2}\bra{m_3} + \textrm{h.c.} \right]  + \frac{\eta^2_t}{2} \rho_{m_4} \left(\frac{\cos^2 \theta + 1}{2}\right) \right],
\end{align}
where $\textrm{h.c}$ represents the Hermitian conjugate of the neighboring term and where we introduced the states $\rho_{m_i}$, defined as:
\begin{align}
    \rho_{m_i} := \Tr_{L_A L_B} \left[\ketbra{m_i}{m_i}\right] \quad\quad i=1,2,3,4.
\end{align}

Straightforward calculations yield explicit expressions for the states $\rho_{m_i}$, which read:
\begin{align}
    \rho_{m_i} &= x_{i,\uparrow\uparrow}\ketbra{\uparrow\uparrow}{\uparrow\uparrow} +x_{i,\uparrow\downarrow}\ketbra{\uparrow\downarrow}{\uparrow\downarrow} + x_{i,\downarrow\uparrow}\ketbra{\downarrow\uparrow}{\downarrow\uparrow} +  x_{i,\downarrow\downarrow} \ketbra{\downarrow\downarrow}{\downarrow\downarrow} \label{rho_m_i}
\end{align}
where the coefficients of $\rho_{m_1}$ are:
\begin{align}
    x_{1,\uparrow\uparrow} &= \left[1 + \sin (2 \alpha ) \cos (\alpha') \right] \left[1 + \sin (2 \beta ) \cos (\beta')\right] = : \tilde{x}_{\uparrow\uparrow}  \label{xupup}\\
    x_{1,\uparrow\downarrow} &=  (1-\eta_t) \left[1 + \sin (2 \alpha ) \cos (\alpha') \right] \left[1 - \sin (2 \beta ) \cos (\beta')\right] =: (1-\eta_t) \, \tilde{x}_{\uparrow\downarrow}  \label{xupdown}\\
    x_{1,\downarrow\uparrow} &= (1-\eta_t) \left[1 - \sin (2 \alpha ) \cos (\alpha') \right] \left[1 + \sin (2 \beta ) \cos (\beta')\right] =: (1-\eta_t)\, \tilde{x}_{\downarrow\uparrow}  \label{xdownup} \\
    x_{1,\downarrow\downarrow} &= (1-\eta_t)^2 \left[1 - \sin (2 \alpha ) \cos (\alpha') \right] \left[1 - \sin (2 \beta ) \cos (\beta')\right] =: (1-\eta_t)^2 \,\tilde{x}_{\downarrow\downarrow}, \label{xdowndown}
\end{align}

the coefficients of $\rho_{m_2}$ are:
\begin{align}
    x_{2,\uparrow\uparrow} &= 0 \\
    x_{2,\uparrow\downarrow} &= \tilde{x}_{\uparrow\downarrow} \\
    x_{2,\downarrow\uparrow} &= 0 \\
    x_{2,\downarrow\downarrow} &= (1-\eta_t)\, \tilde{x}_{\downarrow\downarrow},
\end{align}

the coefficients of $\rho_{m_3}$ are:
\begin{align}
    x_{3,\uparrow\uparrow} &= 0 \\
    x_{3,\uparrow\downarrow} &= 0 \\
    x_{3,\downarrow\uparrow} &=  \tilde{x}_{\downarrow\uparrow} \\
    x_{3,\downarrow\downarrow} &=  (1-\eta_t)\, \tilde{x}_{\downarrow\downarrow},
\end{align}

and the coefficients of $\rho_{m_4}$ are:
\begin{align}
    x_{4,\uparrow\uparrow} &= 0 \\
    x_{4,\uparrow\downarrow} &= 0 \\
    x_{4,\downarrow\uparrow} &=  0 \\
    x_{4,\downarrow\downarrow} &=  \tilde{x}_{\downarrow\downarrow}.
\end{align}
Moreover, we observe that:
\begin{align}
    \Tr_{L_A L_B} \left[e^{-i \varphi}  \ket{m_2}\bra{m_3} + \textrm{h.c.} \right] &= e^{-i \varphi} v \ketbra{\uparrow\downarrow}{\downarrow\uparrow} + e^{i \varphi} v^* \ketbra{\downarrow\uparrow}{\uparrow\downarrow}, \label{extraterm}
\end{align}
where:
\begin{align}
    v= \left[\cos (2 \alpha) + i\sin (2 \alpha )\sin (\alpha')\right] \left[\cos (2 \beta )-i \sin (2 \beta ) \sin (\beta')\right]. \label{v}
\end{align}

By plugging the expressions \eqref{rho_m_i} and \eqref{extraterm} into \eqref{eq: final_state_1_m}, we obtain the following state at the end of step 3. of the BK protocol: 
\begin{align}
\rho_{M_A M_B|\nicefrac{C}{D}\checkmark} 
    &=\frac{1-p_d}{4\Pr(C\checkmark)}\left\lbrace p_d \left[ \tilde{x}_{\uparrow\uparrow} \ketbra{\uparrow\uparrow}{\uparrow\uparrow} +(1-\eta_t)\left( \tilde{x}_{\uparrow\downarrow}\ketbra{\uparrow\downarrow}{\uparrow\downarrow} + \tilde{x}_{\downarrow\uparrow} \ketbra{\downarrow\uparrow}{\downarrow\uparrow}\right)+ (1-\eta_t)^2 \tilde{x}_{\downarrow\downarrow}\ketbra{\downarrow\downarrow}{\downarrow\downarrow})\right] \right.\nonumber\\
    &\quad\left. + \eta_t \left[\frac{\tilde{x}_{\uparrow\downarrow}}{2} \ketbra{\uparrow\downarrow}{\uparrow\downarrow}  + \frac{\tilde{x}_{\downarrow\uparrow}}{2} \ketbra{\downarrow\uparrow}{\downarrow\uparrow} \pm \frac{\cos\theta}{2} (e^{-i\varphi}v \ketbra{\uparrow\downarrow}{\downarrow\uparrow}+e^{i\varphi} v^* \ketbra{\downarrow\uparrow}{\uparrow\downarrow} )\right. \right.\nonumber\\
    &\quad\left.\left. + \left(1-\eta_t\left(\frac{3}{4}-\frac{\cos^2\theta}{4}\right)\right) \tilde{x}_{\downarrow\downarrow} \ketbra{\downarrow\downarrow}{\downarrow\downarrow}\right] \right\rbrace , \label{eq: rho|C-step3-imperfectstateprep}
\end{align}
Note that if one assumes perfect state preparation: $\alpha=\beta=0$, it holds: $\tilde{x}_{\uparrow\uparrow}=\tilde{x}_{\uparrow\downarrow}=\tilde{x}_{\downarrow\uparrow}=\tilde{x}_{\downarrow\downarrow}=v=1$.
\vspace{1ex}

We now proceed with step~4.~of the BK protocol, where we apply the Pauli $X$ gate on both memory qubits. The resulting states of the quantum memories are:
\begin{align} 
\rho_{M_A M_B|\nicefrac{C}{D}\checkmark} 
    &=\frac{1-p_d}{4\Pr(C\checkmark)}\left\lbrace p_d \left[ \tilde{x}_{\uparrow\uparrow} \ketbra{\downarrow\downarrow}{\downarrow\downarrow} +(1-\eta_t)\left( \tilde{x}_{\uparrow\downarrow}\ketbra{\downarrow\uparrow}{\downarrow\uparrow} + \tilde{x}_{\downarrow\uparrow} \ketbra{\uparrow\downarrow}{\uparrow\downarrow}\right)+ (1-\eta_t)^2 \tilde{x}_{\downarrow\downarrow}\ketbra{\uparrow\uparrow}{\uparrow\uparrow})\right] \right.\nonumber\\
    &\quad\left. + \eta_t \left[\frac{\tilde{x}_{\uparrow\downarrow}}{2} \ketbra{\downarrow\uparrow}{\downarrow\uparrow}  + \frac{\tilde{x}_{\downarrow\uparrow}}{2} \ketbra{\uparrow\downarrow}{\uparrow\downarrow} \pm \frac{\cos\theta}{2} (e^{-i\varphi}v \ketbra{\downarrow\uparrow}{\uparrow\downarrow}+e^{i\varphi} v^* \ketbra{\uparrow\downarrow}{\downarrow\uparrow} )\right. \right.\nonumber\\
    &\quad\left.\left. + \left(1-\eta_t\left(\frac{3}{4}-\frac{\cos^2\theta}{4}\right)\right) \tilde{x}_{\downarrow\downarrow} \ketbra{\uparrow\uparrow}{\uparrow\uparrow}\right] \right\rbrace , \label{eq: rho|C-step4-imperfectstateprep}
\end{align}

With step~5., we excite the memories with the $\pi$-pulse, thereby inducing the following transitions for the memory $M_A$:
\begin{align}
    \ket{\uparrow}_{M_A} &\to \ket{\uparrow}_{M_A}\ket{000}_{L_A A A_\perp} \\
    \ket{\downarrow}_{M_A} &\to \ket{\downarrow}_{M_A}\left[\sqrt{\eta_t}\ket{0}_{L_A} e^{i \varphi} (\cos\theta \ket{10}_{A A_\perp} + \sin\theta \ket{01}_{A A_\perp}) + \sqrt{1-\eta_t} \ket{1}_{L_A} \ket{00}_{A A_\perp} \right],
\end{align}
where we included already the global loss $\eta_t$ and the two mode mismatch effects. For the memory $M_B$, we have:
\begin{align}
    \ket{\uparrow}_{M_B} &\to \ket{\uparrow}_{M_B}\ket{000}_{L_B B B_\perp} \\
    \ket{\downarrow}_{M_B} &\to \ket{\downarrow}_{M_B}\left[\sqrt{\eta_t}\ket{0}_{L_B} \ket{10}_{B B_\perp} + \sqrt{1-\eta_t} \ket{1}_{L_B} \ket{00}_{B B_\perp} \right].
\end{align}
With the above transitions, we can compute the resulting state of the memories and the light modes by individually evaluating each term in \eqref{eq: rho|C-step4-imperfectstateprep}. For example, the term $\ket{\uparrow\downarrow}_{M_A M_B}$ is mapped to:
\begin{align}
    \ket{\uparrow\downarrow}\ket{000}_{L_A A A_\perp}\left[\sqrt{\eta_t}\ket{0}_{L_B} \ket{10}_{B B_\perp} + \sqrt{1-\eta_t} \ket{1}_{L_B} \ket{00}_{B B_\perp} \right].
\end{align}

After interfering the the modes $A$, $B$, $A_{\perp}$ and $B_{\perp}$ in the balanced beam splitter, we can compute the resulting memory-light states and obtain:
\begin{align}
    &\tilde{\rho}_{M_A M_B L_A L_B C D C_{\perp} D_{\perp}|\nicefrac{C}{D}\checkmark} =\nonumber\\
    &\left\lbrace p_d\left[\tilde{x}_{\uparrow\uparrow} \ketbra{\downarrow\downarrow}{\downarrow\downarrow} \otimes \ketbra{\tilde{\Phi}}{\tilde{\Phi}} +(1-\eta_t)(\tilde{x}_{\downarrow\uparrow}\ketbra{\uparrow\downarrow}{\uparrow\downarrow} \otimes \ketbra{\tilde{\zeta}}{\tilde{\zeta}} + \tilde{x}_{\uparrow\downarrow}\ketbra{\downarrow\uparrow}{\downarrow\uparrow} \otimes \ketbra{\tilde{\chi}}{\tilde{\chi}}) \right.\right. \nonumber\\
    &\left.\left.\quad + (1-\eta_t)^2 \tilde{x}_{\downarrow\downarrow} \ketbra{\uparrow\uparrow}{\uparrow\uparrow} \otimes \ketbra{000000}{000000}\right] \right.\nonumber\\
    &\quad\left. + \eta_t \left[ \frac{\tilde{x}_{\downarrow\uparrow}}{2}\ketbra{\uparrow\downarrow}{\uparrow\downarrow} \otimes \ketbra{\tilde{\zeta}}{\tilde{\zeta}} + \frac{\tilde{x}_{\uparrow\downarrow}}{2}\ketbra{\downarrow\uparrow}{\downarrow\uparrow} \otimes \ketbra{\tilde{\chi}}{\tilde{\chi}} \right.\right.\nonumber\\
    &\left.\left.\quad \pm \frac{\cos\theta}{2} (e^{i\varphi} v^* \ketbra{\uparrow\downarrow}{\downarrow\uparrow}\otimes\ketbra{\tilde{\zeta}}{\tilde{\chi}} + e^{-i\varphi} v \ketbra{\downarrow\uparrow}{\uparrow\downarrow}\otimes\ketbra{\tilde{\chi}}{\tilde{\zeta}}) \right.\right.\nonumber\\
    &\quad\left.\left.  +  \left(1-\eta_t\left(\frac{3}{4}-\frac{\cos^2\theta}{4}\right)\right) \tilde{x}_{\downarrow\downarrow} \ketbra{\uparrow\uparrow}{\uparrow\uparrow} \otimes \ketbra{000000}{000000}  \right]\right\rbrace (1-p_d) \Big/ 4\Pr(C\checkmark), \label{eq: rho|C-step5}
\end{align}
where we introduced the following pure states:
\begin{align}
    \ket{\tilde{\Phi}}_{L_A L_B C D C_{\perp} D_{\perp}} &= (1-\eta_t) \ket{11}_{L_A L_B} \ket{0000}_{CD C_\perp D_{\perp}} + \sqrt{\eta_t (1-\eta_t)} \frac{e^{i\varphi}\cos\theta \ket{01}_{L_A L_B} + \ket{10}_{L_A L_B}}{\sqrt{2}} \ket{1000}_{CD C_\perp D_{\perp}} \nonumber\\
    &\quad  + \sqrt{\eta_t (1-\eta_t)} \frac{e^{i\varphi}\cos\theta \ket{01}_{L_A L_B} - \ket{10}_{L_A L_B}}{\sqrt{2}} \ket{0100}_{CD C_\perp D_{\perp}} \nonumber\\
    &\quad  + \sqrt{\eta_t (1-\eta_t)} e^{i\varphi}\sin\theta \ket{01}_{L_A L_B} \frac{\ket{0010}_{CD C_\perp D_{\perp}} + \ket{0001}_{CD C_\perp D_{\perp}}}{\sqrt{2}}  \nonumber\\
    &\quad+\eta_t e^{i \varphi} \ket{00}_{L_A L_B} \left(\cos\theta \frac{\ket{2000}- \ket{0200}}{\sqrt{2}} + \sin\theta \frac{\ket{1010} + \ket{1001} -\ket{0110} -\ket{0101}}{2}\right)_{CD C_\perp D_{\perp}}
\end{align}
\begin{align}
    \ket{\tilde{\zeta}}_{L_A L_B C D C_{\perp} D_{\perp}} &= \sqrt{\eta_t}\ket{00}_{L_AL_B} \frac{\ket{1000}_{CD C_\perp D_{\perp}} - \ket{0100}_{CD C_\perp D_{\perp}}}{\sqrt{2}} + \sqrt{1-\eta_t} \ket{01}_{L_A L_B} \ket{0000}_{CD C_\perp D_{\perp}}
\end{align}
\begin{align}
     \ket{\tilde{\chi}}_{L_A L_B C D C_{\perp} D_{\perp}} &= e^{i\varphi}\sqrt{\eta_t} \ket{00}_{L_AL_B} \nonumber\\
     &\quad \left(\cos{\theta} \frac{\ket{1000}_{CD C_\perp D_{\perp}} + \ket{0100}_{CD C_\perp D_{\perp}}}{\sqrt{2}} + \sin{\theta} \frac{\ket{0010}_{CD C_\perp D_{\perp}} + \ket{0001}_{CD C_\perp D_{\perp}}}{\sqrt{2}}\right) \nonumber\\
    &\quad + \sqrt{1-\eta_t}\ket{10}_{L_A L_B} \ket{0000}_{CD C_\perp D_{\perp}}.
\end{align}
\vspace{1ex}
Now, by post-selecting again the state on a click in detector $C$, the memories are left in the following unnormalized state:
\begin{align}
    &\Pr(C\checkmark|\nicefrac{C}{D}\checkmark) \rho_{M_A M_B|C\checkmark, \nicefrac{C}{D}\checkmark} = \nonumber\\
    &\Tr_{L_A L_B C D C_\perp D_{\perp}}\left[(\one_C \otimes \one_{C_\perp }-(1-p_d)\ketbra{00}{00}_{C C_\perp }) \otimes (1-p_d)\ketbra{00}{00}_{D D_\perp}  \tilde{\rho}_{M_A M_B L_A L_B C D C_{\perp} D_{\perp}|\nicefrac{C}{D} \checkmark} \right] \nonumber\\
    &=\left\lbrace p_d^2 (1-\eta_t)^2 (\tilde{x}_{\uparrow\uparrow}\ketbra{\downarrow\downarrow}{\downarrow\downarrow} + \tilde{x}_{\downarrow\uparrow}\ketbra{\uparrow\downarrow}{\uparrow\downarrow} + \tilde{x}_{\uparrow\downarrow}\ketbra{\downarrow\uparrow}{\downarrow\uparrow} +\tilde{x}_{\downarrow\downarrow}\ketbra{\uparrow\uparrow}{\uparrow\uparrow}) \right. \nonumber\\
    &\left.\quad + p_d \eta_t \left[\frac{1-\eta_t}{2}\tilde{x}_{\downarrow\uparrow} \ketbra{\uparrow\downarrow}{\uparrow\downarrow} + \frac{1-\eta_t}{2} \tilde{x}_{\uparrow\downarrow} \ketbra{\downarrow\uparrow}{\downarrow\uparrow} + \left(1-\eta_t\left(\frac{3}{4}-\frac{\cos^2\theta}{4}\right)\right) \tilde{x}_{\downarrow\downarrow} \ketbra{\uparrow\uparrow}{\uparrow\uparrow}\right] \right. \nonumber\\
    &\left.\quad+ p_d \eta_t \left[ \left(1- \eta_t \left(\frac{3}{4} - \frac{\cos^2 \theta}{4}\right)\right) \tilde{x}_{\uparrow\uparrow}\ketbra{\downarrow\downarrow}{\downarrow\downarrow} + \frac{1-\eta_t}{2} \tilde{x}_{\downarrow\uparrow} \ketbra{\uparrow\downarrow}{\uparrow\downarrow} + \frac{1-\eta_t}{2} \tilde{x}_{\uparrow\downarrow} \ketbra{\downarrow\uparrow}{\downarrow\uparrow}\right] \right.\nonumber\\ 
    &\left.\quad  +\frac{\eta^2_t}{2} \left[\frac{\tilde{x}_{\downarrow\uparrow}}{2}\ketbra{\uparrow\downarrow}{\uparrow\downarrow} + \frac{\tilde{x}_{\uparrow\downarrow}}{2} \ketbra{\downarrow\uparrow}{\downarrow\uparrow} \pm  \frac{\cos^2 \theta}{2} \left(v^*\ketbra{\uparrow\downarrow}{\downarrow\uparrow} + v \ketbra{\downarrow\uparrow}{\uparrow\downarrow}\right)\right] \right\rbrace  (1-p_d)^2 \Big/ 4\Pr(C\checkmark) ,
\end{align}
which can be made more compact as follows:
\begin{align} \label{eq: unnormC-imperfectstateprep}
    &\Pr(C\checkmark|\nicefrac{C}{D}\checkmark) \rho_{M_A M_B|C\checkmark, \nicefrac{C}{D}\checkmark} = \left\lbrace p_d^2 (1-\eta_t)^2 (\tilde{x}_{\uparrow\uparrow}\ketbra{\downarrow\downarrow}{\downarrow\downarrow} + \tilde{x}_{\downarrow\uparrow}\ketbra{\uparrow\downarrow}{\uparrow\downarrow} + \tilde{x}_{\uparrow\downarrow}\ketbra{\downarrow\uparrow}{\downarrow\uparrow} +\tilde{x}_{\downarrow\downarrow}\ketbra{\uparrow\uparrow}{\uparrow\uparrow}) \right. \nonumber\\
    &\left.\quad + p_d \eta_t \left[(1-\eta_t) \left(\tilde{x}_{\downarrow\uparrow} \ketbra{\uparrow\downarrow}{\uparrow\downarrow} + \tilde{x}_{\uparrow\downarrow} \ketbra{\downarrow\uparrow}{\downarrow\uparrow} \right) + \left(1-\eta_t\left(\frac{3}{4}-\frac{\cos^2\theta}{4}\right)\right) \left(\tilde{x}_{\downarrow\downarrow} \ketbra{\uparrow\uparrow}{\uparrow\uparrow} + \tilde{x}_{\uparrow\uparrow}\ketbra{\downarrow\downarrow}{\downarrow\downarrow} \right)\right] \right. \nonumber\\
    &\left.\quad  +\frac{\eta^2_t}{2} \left[\frac{\tilde{x}_{\downarrow\uparrow}}{2}\ketbra{\uparrow\downarrow}{\uparrow\downarrow} + \frac{\tilde{x}_{\uparrow\downarrow}}{2} \ketbra{\downarrow\uparrow}{\downarrow\uparrow} \pm  \frac{\cos^2 \theta}{2} \left(v^*\ketbra{\uparrow\downarrow}{\downarrow\uparrow} + v \ketbra{\downarrow\uparrow}{\uparrow\downarrow}\right)\right] \right\rbrace  (1-p_d)^2 \Big/ 4\Pr(C\checkmark) .
\end{align}

The resulting probability of detector $C$ clicking, given that either detector clicked in the first BSM, is given by:
\begin{align} \label{eq: pc_final}
    &\Pr(C\checkmark|\nicefrac{C}{D}\checkmark) = \frac{(1-p_d)^2}{4\Pr(C\checkmark)} \left\lbrace p_d^2 (1-\eta_t)^2 (\tilde{x}_{\uparrow\uparrow} + \tilde{x}_{\downarrow\uparrow} + \tilde{x}_{\uparrow\downarrow} +\tilde{x}_{\downarrow\downarrow}) \right. \nonumber\\
    &\left.\quad + p_d \eta_t \left[(1-\eta_t) \left(\tilde{x}_{\downarrow\uparrow}  + \tilde{x}_{\uparrow\downarrow}  \right) + \left(1-\eta_t\left(\frac{3}{4}-\frac{\cos^2\theta}{4}\right)\right) \left(\tilde{x}_{\downarrow\downarrow} + \tilde{x}_{\uparrow\uparrow}\right)\right] +\frac{\eta^2_t}{4}(\tilde{x}_{\downarrow\uparrow} + \tilde{x}_{\uparrow\downarrow}) \right\rbrace 
\end{align}

Similarly, we post-select the state in \eqref{eq: rho|C-step5} on a click in detector $D$, thus leaving the memories in the following unnormalized state:
\begin{align}  \label{eq: unnormD-imperfectstateprep}
    &\Pr(D\checkmark|\nicefrac{C}{D}\checkmark) \rho_{M_A M_B|D\checkmark, \nicefrac{C}{D}\checkmark} = \left\lbrace p_d^2 (1-\eta_t)^2 (\tilde{x}_{\uparrow\uparrow}\ketbra{\downarrow\downarrow}{\downarrow\downarrow} + \tilde{x}_{\downarrow\uparrow}\ketbra{\uparrow\downarrow}{\uparrow\downarrow} + \tilde{x}_{\uparrow\downarrow}\ketbra{\downarrow\uparrow}{\downarrow\uparrow} +\tilde{x}_{\downarrow\downarrow}\ketbra{\uparrow\uparrow}{\uparrow\uparrow}) \right. \nonumber\\
    &\left.\quad + p_d \eta_t \left[(1-\eta_t) \left(\tilde{x}_{\downarrow\uparrow} \ketbra{\uparrow\downarrow}{\uparrow\downarrow} + \tilde{x}_{\uparrow\downarrow} \ketbra{\downarrow\uparrow}{\downarrow\uparrow} \right) + \left(1-\eta_t\left(\frac{3}{4}-\frac{\cos^2\theta}{4}\right)\right) \left(\tilde{x}_{\downarrow\downarrow} \ketbra{\uparrow\uparrow}{\uparrow\uparrow} + \tilde{x}_{\uparrow\uparrow}\ketbra{\downarrow\downarrow}{\downarrow\downarrow} \right)\right] \right. \nonumber\\
    &\left.\quad  +\frac{\eta^2_t}{2} \left[\frac{\tilde{x}_{\downarrow\uparrow}}{2}\ketbra{\uparrow\downarrow}{\uparrow\downarrow} + \frac{\tilde{x}_{\uparrow\downarrow}}{2} \ketbra{\downarrow\uparrow}{\downarrow\uparrow} \mp  \frac{\cos^2 \theta}{2} \left(v^*\ketbra{\uparrow\downarrow}{\downarrow\uparrow} + v \ketbra{\downarrow\uparrow}{\uparrow\downarrow}\right)\right] \right\rbrace  (1-p_d)^2 \Big/ 4\Pr(C\checkmark) ,
\end{align}
with probability:
\begin{align}
    \Pr(D\checkmark|\nicefrac{C}{D}\checkmark) = \Pr(C\checkmark|\nicefrac{C}{D}\checkmark).
\end{align}
By comparing \eqref{eq: unnormC-imperfectstateprep} and \eqref{eq: unnormD-imperfectstateprep}, we observe that if the same detector clicked in both excitation rounds, then the state of the memories contains the $\cos^2 \theta/2$ contribution with a plus sign. Vice-versa, if two different detectors clicked, the two memories are described by a state containing the $\cos^2 \theta/2$ contribution with a minus sign. The final step of the BK protocol applies a Pauli $Z$ on the second memory, in case two different detectors clicked in the two excitation rounds. By doing so, we map the two resulting states, conditioned on equal or different detector clicks, into the same state, with the plus sign.

In summary, the normalized state of the two quantum memories at the end of the BK protocol, regardless of the click pattern in the two excitation rounds, reads:
\begin{align} \label{finalstate-BKprotocol-imperfectstateprep}
    &\rho_{M_A M_B|\checkmark, \checkmark} =  \left\lbrace p_d^2 (1-\eta_t)^2 (\tilde{x}_{\uparrow\uparrow}\ketbra{\downarrow\downarrow}{\downarrow\downarrow} + \tilde{x}_{\downarrow\uparrow}\ketbra{\uparrow\downarrow}{\uparrow\downarrow} + \tilde{x}_{\uparrow\downarrow}\ketbra{\downarrow\uparrow}{\downarrow\uparrow} +\tilde{x}_{\downarrow\downarrow}\ketbra{\uparrow\uparrow}{\uparrow\uparrow}) \right. \nonumber\\
    &\left.\quad + p_d \eta_t \left[(1-\eta_t) \left(\tilde{x}_{\downarrow\uparrow} \ketbra{\uparrow\downarrow}{\uparrow\downarrow} + \tilde{x}_{\uparrow\downarrow} \ketbra{\downarrow\uparrow}{\downarrow\uparrow} \right) + \left(1-\eta_t\left(\frac{3}{4}-\frac{\cos^2\theta}{4}\right)\right) \left(\tilde{x}_{\downarrow\downarrow} \ketbra{\uparrow\uparrow}{\uparrow\uparrow} + \tilde{x}_{\uparrow\uparrow}\ketbra{\downarrow\downarrow}{\downarrow\downarrow} \right)\right] \right. \nonumber\\
    &\left.\quad  +\frac{\eta^2_t}{2} \left[\frac{\tilde{x}_{\downarrow\uparrow}}{2}\ketbra{\uparrow\downarrow}{\uparrow\downarrow} + \frac{\tilde{x}_{\uparrow\downarrow}}{2} \ketbra{\downarrow\uparrow}{\downarrow\uparrow} +  \frac{\cos^2 \theta}{2} \left(v^*\ketbra{\uparrow\downarrow}{\downarrow\uparrow} + v \ketbra{\downarrow\uparrow}{\uparrow\downarrow}\right)\right] \right\rbrace \Bigg/ \nonumber\\
    &\quad\left\lbrace p_d^2 (1-\eta_t)^2 (\tilde{x}_{\uparrow\uparrow} + \tilde{x}_{\downarrow\uparrow} + \tilde{x}_{\uparrow\downarrow} +\tilde{x}_{\downarrow\downarrow}) + p_d \eta_t \left[(1-\eta_t) \left(\tilde{x}_{\downarrow\uparrow}  + \tilde{x}_{\uparrow\downarrow}  \right) + \left(1-\eta_t\left(\frac{3}{4}-\frac{\cos^2\theta}{4}\right)\right) \left(\tilde{x}_{\downarrow\downarrow} + \tilde{x}_{\uparrow\uparrow}\right)\right] \right.\nonumber\\
    &\left.\quad +\frac{\eta^2_t}{4}(\tilde{x}_{\downarrow\uparrow} + \tilde{x}_{\uparrow\downarrow}) \right\rbrace.
\end{align}
and the overall success probability of the protocol is given by:
\begin{align}
    \Pr(\textrm{success}) &= 4 \Pr(C\checkmark|\nicefrac{C}{D}\checkmark) \Pr(C \checkmark) \nonumber\\
    &=(1-p_d)^2 \left\lbrace p_d^2 (1-\eta_t)^2 (\tilde{x}_{\uparrow\uparrow} + \tilde{x}_{\downarrow\uparrow} + \tilde{x}_{\uparrow\downarrow} +\tilde{x}_{\downarrow\downarrow}) \right. \nonumber\\
    &\left.\quad + p_d \eta_t \left[(1-\eta_t) \left(\tilde{x}_{\downarrow\uparrow}  + \tilde{x}_{\uparrow\downarrow}  \right) + \left(1-\eta_t\left(\frac{3}{4}-\frac{\cos^2\theta}{4}\right)\right) \left(\tilde{x}_{\downarrow\downarrow} + \tilde{x}_{\uparrow\uparrow}\right)\right] +\frac{\eta^2_t}{4}(\tilde{x}_{\downarrow\uparrow} + \tilde{x}_{\uparrow\downarrow}) \right\rbrace . \label{probsuccess-BKprotocol-imperfectstateprep}
\end{align}
The fidelity of the final state of the memories, given in \eqref{finalstate-BKprotocol-imperfectstateprep}, with the ideal output, i.e. the Bell state $\ket{\Psi^+}$, is given by:
\begin{align} \label{fidelity-BKprotocol-imperfectstateprep}
    \mathcal{F} &= \braket{\Psi^+|\rho_{M_A M_B|\checkmark \checkmark}|\Psi^+} \nonumber\\
    &= \left\lbrace p_d^2 (1-\eta_t)^2 ( \tilde{x}_{\downarrow\uparrow} + \tilde{x}_{\uparrow\downarrow})/2  + p_d \eta_t (1-\eta_t) \left(\tilde{x}_{\downarrow\uparrow}  + \tilde{x}_{\uparrow\downarrow}  \right)/2   +\frac{\eta^2_t}{8} \left(\tilde{x}_{\downarrow\uparrow} + \tilde{x}_{\uparrow\downarrow}  +  2\Re(v) \cos^2 \theta  \right) \right\rbrace (1-p_d)^2 \nonumber\\
    &\quad \left(\Pr(\textrm{success})\right)^{-1} \nonumber\\
    &=\left\lbrace \frac{\tilde{x}_{\downarrow\uparrow} + \tilde{x}_{\uparrow\downarrow}}{2} \left[p_d(1-\eta_t) + \frac{\eta_t}{2}\right]^2 + \frac{\eta_t^2}{4} \Re(v) \cos^2 \theta \right\rbrace (1-p_d)^2 \Big/ \Pr(\textrm{success}).
\end{align}
The explicit expressions of the parameters $\tilde{x}_{\uparrow\uparrow}$,$\tilde{x}_{\uparrow\downarrow}$, $\tilde{x}_{\downarrow\uparrow}$, $\tilde{x}_{\downarrow\downarrow}$, and $v$ appearing in the formulas for the state, the success probability and the fidelity are given in \eqref{xupup}, \eqref{xupdown}, \eqref{xdownup}, \eqref{xdowndown}, and \eqref{v}, respectively.

\section{Photon indistinguishability and
visibility of the two-photon interference} \label{sec: photon_ind}
In this Appendix the visibility of the two-photon interference, an important experimental parameter used to quantify the degree of indistinguishability of two photons, is related to our error parameters and in particular with $\theta$.

The visibility of two-photon interference can be obtained directly from the second-order autocorrelation function $g^{(2)}(dt)$. This function is proportional to the histogram of all coincidences obtained between photons detected on the two detectors after the beam splitter, where $dt = t_C - t_D$, and $t_C$ and $t_D$ are the arrival times of the photons detected by $C$ and $D$, respectively. In particular, for fully distinguishable photon the functions will have three peaks. Two are at $dt = \pm T$, where $T$ is the length of an excitation round. They represent both photons arriving at the same time in the same detector. These peaks do not depend on the distinguishability of the photons. The third peak, instead, is at $dt = 0$. It represents the two photons arriving at the same time in both detectors. It is half of the other two peaks when the photons are fully distinguishable (orthogonal polarization, or $\theta = \frac{\pi}{2}$), and is smaller the less the photons are distinguishable, until it disappears completely for fully indistinguishable photons.

The visibility $V(dt)$ of the interference at $dt$ is defined in terms of the second-order autocorrelation function as follows:

\begin{equation} \label{eq: indistinguishability}
    V(dt) = \frac{g_\perp^{(2)}(dt) - g^{(2)}(dt)}{g_\perp^{(2)}(dt)},
\end{equation}

where $g_\perp^{(2)}(dt)$ is the function in the fully distinguishable case. Normally, the photon indistinguishability $I$ is defined as the visibility of the two-photon interference at zero time delay ($dt=0$), i.e., $I=V(0)$.

By renormalizing the autocorrelation function by the side peaks, one obtains the function $\tilde{g}_\perp^{(2)}(0) = \frac{1}{2}$. On the other hand, $\tilde{g}^{(2)}(0)$ can be obtained from the probability of the two detectors triggering simultaneously when no photons are lost. From \eqref{eq: psi_3_with_m} this can easily be obtained as $\tilde{g}^{(2)}(0)=\frac{\sin(\theta)^2}{2}$. Thus, we obtain:
\begin{equation} \label{eq: indistinguishability_from_theta}
    I = V(0) = \frac{\frac{1}{2}-\frac{\sin^2(\theta)}{2}}{\frac{1}{2}} = \cos^2(\theta),
\end{equation}
which directly relates the photon indistinguishability $I$  with the parameter $\theta$ modeling the imperfect mode matching.

\end{document}